\documentclass[10pt,a4paper]{article}
\usepackage{graphpap,epsfig,amssymb,graphicx,textcomp}
\newcommand{\mathsym}[1]{{}}

\begin{document}
\begin{center}
 \large{\bf Transient Chaos Generates Small Chimeras}
\end{center}

\vskip 1cm

\begin{center}{\it Amitava Banerjee$^{1}$ and Debopriya Sikder$^{2}$}\\
\vskip 0.2 cm
{\it Department of Physics, Presidency University,}\\
{\it 86/1 College street, Kolkata-700073, INDIA}\\
\vskip 1 cm
{1. E-mail: amitava8196@gmail.com}\\
{2. E-mail: debopriyasikder@gmail.com}\end{center}

\begin{abstract}
 While the chimera states themselves are usually believed to be chaotic transients, the involvement of chaos behind their self-organization is not properly distinguished or studied.  In this work, we demonstrate that small chimeras in the local flux dynamics of an array of magnetically coupled superconducting quantum interference devices (SQUIDs) driven by an external field are born through transiently chaotic dynamics. We deduce analytic expressions for small chimeras and synchronous states which correspond to nonchaotic attractors in the model. We also numerically study the bifurcations underlying the multistability responsible for their generation. Transient chaos manifests itself in the short term flux oscillations with erratically fluctuating amplitudes, exponential escape time distribution and irregular dependence of the escape time to initial conditions. We classify the small chimera states in terms of the position of the non-synchronized member and numerically construct their basin of attraction. The basin is shown to possess an interesting structure consisting of both ordered and fractal parts, which again can be attributed to transient chaos. 
\end{abstract}
\vskip 0.5 cm

\vskip 1cm
\noindent {\bf I. INTRODUCTION}
\vskip 0.5 cm
  The intriguing coexistence of synchronous and asynchronous groups in a collection of non-locally interacting identical oscillators, known as \textit{chimera} or \textit{multichimera} states \cite{panaggio,pan1}, are as enigmatic as the Greek mythological creature with incongruous body parts that they are named after. Following the 2002 discovery of these states in the paradigmatic Kuramoto-Sakaguchi model \cite{kur1}, they have been speculated to be related to a diverse range of contexts. Examples include neural dynamics during unihemispheric sleeps of some birds and mammals \cite{panaggio,ma1a,ma1b,ma1c} and the onset of epileptic seizures in human beings \cite{ral1a}, fatal loss of coordination between human cardiac cells \cite{panaggio} to the laminar-turbulent coexistence in the Couette flow \cite{dwi1a}. Apart from these, chimeras are also numerically and experimentally found in diverse dynamical systems \cite{panaggio} ranging from mechanical \cite{mar1a,mar1b,mar1c,mar1d,mar1e}, spin-torque \cite{zak1} and chemical oscillators \cite{panaggio,tin1a,tin1b,tin1c,tin1d}, certain neuron models \cite{dib1a,dib1b,dib1c,dib1d,dib1e,dib1f} to coupled chaotic discrete maps \cite{hag1a,hag1b,hag1c}. A relatively recent addition, which will be the subject of our study in the later sections, is the dynamics of magnetic fluxes threading superconducting quantum interference devices (SQUIDs) in a quantum metamaterial \cite{laz1a,laz1b,laz1c}, where chimera and multichimera states, predicted from numerical simulations, are presently awaiting experimental verification. Such ubiquity of chimeras, however, does not overshadow their mystery. Far from that, as it turns out \cite{panaggio,onew1}, the morphology of the chimera states are as thought-provoking as are the conditions for their emergence, destruction and their dynamical properties, bifurcations and stability. In fact, the apparently irregular oscillations of the asynchronous oscillators in a chimera state render the system chaotic. This chaotic behavior of chimera states has been characterized by studies on Lyapunov spectra for varied interaction ranges of oscillators and system sizes \cite{wol1,wol2,olmi1}, sensitivity to initial conditions \cite{omel1} as well as typical bifurcation scenarios and Poincare sections \cite{dud1}. By nature, this chaos can be spatial \cite{wol1}, temporal \cite{omel1} or a subtle coexistence of both \cite{dud1}. But it has also been shown that in finite size systems, chimera states have a finite lifetime \cite{tin1a,wol1,olmi1,loo1}, after which they collapse to a globally synchronous state, which implies that these states are examples of transiently chaotic states. 
  
Transient chaos is a feature of certain dissipative systems, where the system evolves erratically for a finite time (called the `escape time') due to the presence of non-attracting chaotic sets and finally settles down to one of the available (chaotic or regular) attractors which then determines the asymptotic dynamics \cite{tel1,tel2}. Such transiently chaotic systems typically exhibit positive finite-time Lyapunov exponents \cite{tel1,tel2}, exponential distributions of escape times \cite{tel1,tel2}, irregular dependence of the escape time on initial conditions \cite{tel1,tel2} and fractal basins of attraction \cite{tel1,tel2}. There are ample evidences of real systems exhibiting such behavior \cite{tel1,tel2}, e.g., chemical reactions \cite{tel1}, NMR voltage signals \cite{tel1,jan1} and driven pendulum \cite{tel1} and the motion of particles of a fluid advecting in the wake of a cylinder \cite{tel1,aref1} to mention few. In this regard, while the chimera states themselves are established as transiently chaotic dynamical states, there are no detailed studies relating the generation of chimera states to transient chaos. In this work, we have drawn that connection explicitly for the case of small chimeras. While it is a common experience from numerical simulations that chimeras are formed after some initial transient dynamics of the system which is typically initiated with a random configuration, there is no existing rigorous study of such dynamics. This is the problem we address in the current work and we explicitly show that such dynamics is a chaotic one over a finite time scale, which ultimately gives way to chimeras or synchronous states.

The system we shall focus on an artificially fabricated medium made of a lattice of magnetically coupled identical SQUIDs, each of which is basically a superconducting ring interrupted by a single Josephson junction. The material as a whole interacts with an incident electromagnetic field in some well-defined manner which, owing to the high nonlinearity and multistability \cite{jung2a,jung2b,jung2c,jung2d} of the individual SQUIDs, can be tuned by varying experimentally accessible parameters of the system. The chimeras predicted in the flux oscillations of this material also show some interesting unique features, which make them distinct from most of those found in other dynamical models. Firstly, owing to the high degree of multistability of the individual SQUIDs, the chimera states are very sensitive to initial conditions and model parameter values \cite{laz1a,laz1b,laz1c}. Very slight changes can result in a state with a different amount of synchronization, which may not be a chimera at all. The chimera states here also exhibit the rare characteristic that they are seen to appear in even nearest-neighbor-coupled SQUIDs \cite{laz1b}, where in most \cite{panaggio} (though not all \cite{dib1a,dib1b,dib1c,seth1}) other cases, they emerge in systems with long-range but non-global interaction. In the morphology also, these chimeras are peculiar; because they maintain one order of magnitude difference between the amplitudes of oscillations in synchronous and asynchronous groups, with the former usually having lower amplitudes of oscillation of its members \cite{laz1a}. Apart from these reasons, the study of chimeras in this system is further encouraged by the fact that the fabrication of such materials has been recently possible with Hypres Nb/AlO$_x$/Nb junction process on silicon substrates (with the SQUID superconducting transition temperature at $T_{c}=9.2$K) \cite{jung2d,zha1,dai1}, allowing for experimental probes to their peculiar dynamics. Thus, it is worthwhile to numerically investigate the nature of chimera states in more detail for this system.
 
 In this work, we first systematically constructed the numerical phase diagrams for the various collective states of the system to identify the parameter regime for chimera states and understand the mechanism behind their emergence. For studies on transient chaos, we mainly focus on small system sizes. There we show that while the chimeras and synchronized states correspond to regular attractors whose analytic expressions we have derived, the route of the system towards them involves chaotic dynamics on finite time scales. A numerically efficient distinction between the initial chaotic dynamics and the asymptotic dynamics near the attractors in this system is facilitated by the fact that the latter dynamics is characterized by a steady amplitude of oscillation. Thus we have used the temporal fluctuations of oscillation amplitudes to determine the escape time for a specific initial condition. For varied initial conditions, the escape times vary erratically, with their distribution shown to be an exponential one. The final chimera states are distinguished by the position of the non-synchronized oscillator in the lattice. We plot the basins of attraction for each of them, which are shown to be fractal with additional special properties. Thus we obtain all the major signatures of transient chaos in the dynamics preceding the emergence of chimeras, namely, the short term flux oscillations with erratically fluctuating amplitudes, exponential escape time distribution, irregular dependence of the escape time to initial conditions and fractal basins of attraction of the attractors. Thus we establish that chimeras in a small-sized metamaterial (`small chimeras') are born through transient chaos. In future, this study can be generalized to other models and larger lattice sizes and the universality of these results be tested.
 
 This study may turn out to be the key to understanding the mechanism of self-organization of chimeras in larger systems where the chimera state may itself be a chaotic transient \cite{wol1,omel1}. In those systems, if we can quantitatively differentiate between the dynamics preceding a fully developed chimera state and the chaotic dynamics during the evolution of a fully developed chimera state, then we may try to find signatures of transient chaos in the former part as well. If we succeed, we can compare the natures of transient chaos in those two parts of dynamics and establish the chimera-sustaining systems as unique chaotic systems exhibiting sequentially transient dynamics and search for the underlying attracting and nonattracting invariant sets. In fact, in that respect, the typical observation for a chimera state that the position of the synchronous and asynchronous parts in the lattice is strongly dependent on the initial condition \cite{kur1,omel1} is encouraging enough to probe the involvement of transient chaos. This observation seems to be in tune with our findings that the basins of attraction of small chimeras are fractal ones, which also lead to a loss of predictability of the system's final state in terms of the position of the non-synchronized oscillator. From more practical points of view, understanding how the chimera states emerge may lead to their efficient control. More importantly, we may be able to predict, delay or even inhibit their emergence altogether in cases where chimera states and partial synchronization are unwanted, for example, in epileptic seizures \cite{ral1a}, and power grid desynchronization events \cite{panaggio}. In summary, this study, directly relating chaos and birth of chimeras, provides a useful understanding of the mechanisms at play behind the genesis of one of the most important and ubiquitous collective dynamical states. 

The organization of this work is as follows: section II provides an introduction to the model, gives examples of a number of possible dynamical states and explains the phase diagrams of the model, so that we are able to characterize the system completely and choose the pertinent parameter regimes for small chimeras. The next section is devoted to analytic and numerical studies on smallest chimera states: their formation, classification, dynamics, basins of attraction and multiple manifestations of transient chaotic dynamics in them. Finally, section IV summarizes our findings and mentions few directions of future study.
\vskip 1cm
\noindent {\bf II. THE MODEL, COLLECTIVE STATES AND PHASE DIAGRAMS}

\vskip 0.5 cm

\noindent {\scriptsize {\bf A THE MODEL EQUATIONS AND COLLECTIVE STATES}}
\vskip 0.5 cm
In this section, we introduce the mathematical model of the metamaterial which we shall use below. The material is composed of a one-dimensional linear array of equally spaced N identical SQUIDs driven by an external magnetic field. The dynamical variables are the supercurrents $I_{n} (n=1,...,N)$ in the individual SQUID rings and magnetic fluxes $\phi _{n}(n=1,...,N)$ threading them (Fig. 1). A current flowing through a SQUID ring generates a magnetic field around it, which contributes to the fluxes enclosed by the other rings and thus the SQUIDs engage in an effective long-range dipole-dipole interaction. The flux and current in each SQUID are further related and can be described by the the resistively and capacitively shunted junction (RCSJ) SQUID model \cite{likh1}. These two facts enable one to write a set of coupled dimensionless equations for the fluxes only as \cite{laz1a}
\begin{equation}
 \ddot{\phi_{n}} +\gamma\dot{\phi_{n}}+\beta sin(2\pi \phi_{n})=\sum_{m=1}^{N}(\hat\Lambda ^{-1})_{nm}(\phi_{e}-\phi_{m})
\end{equation} 
where $n=1,2..,N$, and $\phi_{e}(t)=\phi_{ac}cos\Omega t$ is the flux enclosed by each ring due to an external time-dependent driving magnetic field and the coupling matrix for the inverse-cubic interaction between the SQUIDs is defined as
\begin{eqnarray}
\label{04}
  {{\hat\Lambda}} = \left\{ \begin{array}{ll}
        1, & \mbox{if $m= n$};\\
        \lambda_0 \, |m-n|^{-3}, & \mbox{if $m\neq n$}.\end{array} \right.   
\end{eqnarray}
Here $\lambda_0$ is the coupling coefficient between nearest neighboring SQUIDs which depends on the lattice constant and other geometrical factors of the metamaterial, $\beta$ is a parameter determining whether a SQUID is hysteretic or not and $\gamma$ is the loss coefficient. One may consult Refs. \cite{laz1a} for the derivation of Eq. (1) and the normalization units of various quantities used in those equations.

In numerical simulations of Eqs. (1) and (2), we initialize the fluxes with values randomly and uniformly chosen from the specified symmetric interval $[-\phi _{R}/2,\phi _{R}/2]$ and also choose the initial flux velocity to be zero all the fluxes, i.e., $\dot{\phi_{n}}(0)=0$ $\forall n$. In actual experiments, as suggested in \cite{laz1a}, the random flux configuration can be achieved by thermal quenching, where the whole set-up is cooled through the superconducting transition temperature $T_{c}$ of the SQUIDs. We numerically solve Eqs. (1) and (2) using the fourth order Runge-Kutta method with a constant step size of $dt=0.02$ unit as in Ref. \cite{laz1a}. The solutions show that, starting from a random flux configuration on the lattice, one usually obtains a number of different ordered states, e.g., (1) synchronous, where all the fluxes coherently oscillate as a mass, (2) cluster, where also the oscillations are mostly synchronous except for a few fluxes at scattered positions throughout the lattice, and (3) chimera and multichimera, where the fluxes spontaneously separate into one or more synchronous groups separated by asynchronous flux groups of similar sizes. Fig. 2 illustrates the diversity of these collective spatiotemporal pattterns.   
\vskip 0.5 cm

\noindent {\scriptsize {\bf B ORDER PARAMETERS AND PHASE DIAGRAMS}}
\vskip 0.5 cm
In order to have a complete idea about the behavior of the system and possible dynamical states in various parameter ranges, we need to identify the collective states of the system. In this section, we identify the parameter regimes corresponding to chimera, multichimera and synchronized states and elucidate the transitions between them, so that we gain an insight into how chimera states are formed and broken. For the necessary systematic identification and statistical characterization of the various collective modes of the system mentioned above, we will use three order parameters. They have been successfully used in earlier works to detect and characterize chimera states in varied range of systems, e.g., from coupled Kuramoto \cite{ban1}, and Lorentz oscillators \cite{gopal1} to Mackey-Glass delayed system \cite{gopal1} and Hidmarsh-Rose neurons \cite{gopal2}. The first of these measures is the traditionally used complex Kuramoto order parameter \cite{acebron1} $R(t)$. At a particular instant of time, this is calculated from a particular flux configuration using the following equation,
\begin{equation}
R(t)=\frac{1}{N}\sum_{n=1}^{N}e^{2\pi i \phi_{n}(t)}.
\end{equation}
The magnitude of this quantity, averaged over a sufficiently large time interval during equilibrium (occurring typically after 5000 time units for $N=64$) encodes the amount of global coherence in the system. In fact, $|R|$ is $0$ for asynchronous state, $1$ for globally synchronous state and increases monotonically in between. Thus, for chimera, multichimera and cluster states, the system is only partially synchronized, so $|R|$ takes some intermediate value between $0$ and $1$. However, since Eq. (3) shows that $|R|$ is unchanged under arbitrary permutations of the fluxes $\phi_{i}$s, most of which obviously destroy their local correlations and chimera order, hence an intermediate value of $|R|$ does not necessarily imply a chimera. To overcome this difficulty, two new order parameters were devised in earlier works \cite{gopal1,gopal2}, which are capable of detecting and distinguishing chimera and multichimera states by calculating the discontinuity of local order in a given flux configuration. These are called strength of incoherence (SI) $S$ and discontinuity measure (DM) $\eta$. To calculate them, as in the earlier works \cite{gopal1,gopal2}, we start by defining $z_i=\phi_{i}-\phi _{i+1}$. Next, we bin the total $N$ fluxes into $N_g$ bins of equal sizes $N_{b}=N/N_{g}$ and calculate the time-averaged variance of $z_i$ for each bin by the usual equation:
\begin{equation}
\sigma_k= \left\langle \sqrt{{\sum_{j=N_b(k-1)+1}^{N_bk}(z_j-{\bar z})^2} \over {N_b}}\right\rangle_{t},
\end{equation}
with $k=1,2,...,N_{g}$ being the bin index. This quantity measures the fluctuation of fluxes in a single bin over the global mean $\bar z=\frac{\Sigma_{i=1}^{N}z_{i}}{N}$. To identify the bins with a relatively lower fluctuation of $z$ (corresponding to an enhanced local coherence), one defines a variable $\lambda$ for each bin by the equation \cite{gopal1,gopal2}
\begin{equation}
\lambda_k=\Theta (\delta - \sigma_k)
\end{equation}
which is $1$ for a bin having variance $\sigma$ lower than a sufficiently small threshold $\delta$ and $0$ for other bins. Summing up this binary measure of local order over the bins, we can then define the strength of incoherence $S$ as \cite{gopal1,gopal2}
\begin{equation}
S = 1 - {{\sum_{k=1}^{k=N_g} \lambda_k} \over {N_g}},
\end{equation}
 so that, for complete flux synchrony $S=0$, for complete asynchrony $S=1$ and $0<S<1$ implies chimera or multichimera states. To further differentiate between the last two states, we count the number of discontinuities separating coherent and incoherent regions in the flux profile as characterized by different values of $\lambda$. Thus the discontinuity measure is defined by summing up the differences between $\lambda $ values of neighboring bins \cite{gopal1,gopal2} using the following equation
\begin{equation}
\eta = \frac{\left(\sum_{k=1}^{k=N_g} |\lambda_k - \lambda_{k+1}|\right)}{2}
\end{equation}
along with the periodic boundary condition $\lambda_{N_{g}+1}=\lambda_{1}$. For both completely synchronous and completely random flux configurations, where the flux profile shows no discontinuity in the coherence profile and thus $\lambda $ is same for all the bins, $\eta$ takes the value 0. However, it is 1 for chimera state and a larger integer for multichimera states since both of them involve coexistence of domains with different amounts of coherence. Apart from these states, we also encountered cluster states. They are, however, harder to distinguish by these measures, and one often has to resort to more direct (often manual) ways \cite{gopal1} for their detection. In our case, we identified them by the fact that unlike chimeras, they do not show significant lowering of $|R|$ from 1, but the SI and DM values do not correspond to synchronous states. Using this observation alongwith the three order parameters, we now have a computationally efficient way of systematically identifying various collective states from the dynamics of the flux profiles in the system.

To deduce the nature of collective states at different parameter values we plot the phase diagrams in $\beta-\gamma$ plane for the three different order parameters $|R|$, SI and DM in Fig. 3. These show, for sufficiently low values of the damping parameter $\gamma $, the system is in an asynchronous state for any value of $\beta$. This can be explained with our observation in the next section that the time required for the system to reach equilibrium increases when the damping parameter is lowered and thus for smaller $\gamma $, steady states such as chimeras and synchronous states are not formed within a reasonable time. For higher $\gamma $ values, as is seen from Fig. 3, the system follows a specific path towards its journey to synchronization through the growth of discontinuity in the system as revealed by the measure $\eta $ (Fig. 3c). Fig. 3 shows, for sufficiently low values of $\beta$, the system is in chimera states. If $\beta $ is increased further, discontinuities are introduced within the chimera states. The chimera gets fragmented and synchronous parts appear within its asynchronous portions and the system moves towards multichimera states as indicated by the increasing DM values. On increasing $\beta $ more, the number of discontinuities in the flux profile increases further and system goes into cluster states with a large number of synchronous portions. Finally those synchronous parts merge together, reducing the number of discontinuities and leading the system to a completely synchronous state. This route shows how increasing the SQUID hysteresis parameter $\beta$ affects the chimera state and ultimately destroys it. This transition, inferred from the amount of discontinuities in the flux profile as measured by $\eta $, is also confirmed and visualized from the results of direct simulation which we plotted in Fig. 2b-2f. In the next section, we shall reconcile this phenomena with the bifurcation diagram of the system with respect to $\beta $ where we shall relate the formation of these discontinuities to the distribution of the fluxes among multiple possible solutions when the system becomes multistable for certain ranges of $\beta $. Thus, the phase diagrams studied here will guide us to find the suitable parameter regimes to look for chimera states with a smaller system size. Furthermore, as we showed, the phase diagrams clarify the dependence of the formation and annihilation of chimeras on the hysteresis parameter. This will complement our study on similar aspects in smaller systems with the bifurcation diagram discussed in the next section.

Since our system of SQUIDs is known to exhibit pronounced multistability \cite{laz1a,laz1b,laz1c}, for consistency we used the same random initial flux configuration for all the points in the phase diagrams. To complement this, we have studied the multistability of the system explicitly to check the robustness of the phase diagrams. For that purpose, histograms of equilibrium values of $|R|$ are plotted for different initial flux configurations (Fig. 4). The histograms show that, away from the various phase boundaries in Fig. 3c, the system is monostable and thus the prediction of various collective states distinguished by $\eta$ in Fig. 3c and hence the mentioned route towards synchronization is statistically robust. only near the phase boundaries the system can exhibit multistability manifested by a wide distribution of the possible equilibrium values of $|R|$ in Fig 4. In this respect, we note that the histograms of the statistical order parameter $|R|$ will fail to detect another kind of multistable behavior if two microscopically distinct states yield identical values of the order parameter. For example, two chimera states with different lattice portions synchronized can have the same $|R|$ value but are distinguishable. For small system size, this issue is studied in detail in the next section. 
\vskip 1cm
\noindent {\bf III. SMALLEST CHIMERA STATES AND TRANSIENT CHAOS}
\vskip 0.5 cm

\noindent {\scriptsize {\bf A CLASSIFICATION AND DYNAMICS OF POSSIBLE SMALL CHIMERA STATES}}
\vskip 0.5 cm
After the discovery and thorough analytic and experimental study of chimera states in a wide variety of systems, there is now a recent interest in realizing these (or at least similar) states in systems of minimum possible size. In this respect one usually adopts the notion of the so-called \textit{weak chimera} states, defined by Ashwin
and Burylko \cite{ash1,ash2} as the states where two or more individuals of a system are frequency synchronized and the rest (one or more) oscillate with a different frequency than the synchronized group. Weak chimeras of this type have been numerically demonstrated in systems of three, four or sometimes a few more identical units for the cases of inertial Kuramoto model \cite{yur1}, Hansel-Mato-Meunier model \cite{ash1} and Lang-Kobayashi-type model of delay-coupled lasers \cite{and2}; and experimentally shown in coupled pendula \cite{wo1} and laser systems \cite{rajar1}. In this section, we shall show that the system of SQUIDS modeled by Eq. (1) with $N=3$ adds to this list. In fact, evidently the system of 3 SQUIDs can be completely synchronized, or they can have chimera states, which may then be distinguished and classified by the position of the non-synchronized SQUID in the lattice. In that respect, Fig. 5 shows that we can obtain all of these possibilities from simulations of Eq. (1) with various initial conditions and parameter values. The plots also show that while the chimeras, cluster states and synchronous states arising in larger systems (as those shown in Fig. 2 and Ref. \cite{laz1a}) typically exhibit larger oscillation amplitudes for asynchronous fluxes and smaller oscillation amplitudes for synchronous fluxes, this is not necessarily the case for small chimeras. Indeed, as we found through direct simulations, large amplitude (magnitude around 1) and globally synchronous oscillation for $\phi _{ac}=0.01$ can occur upto a lattice size of $N=10$. This shows that smaller metamaterial lattices do sustain coherent oscillations with amplitudes much larger than that of the driving flux. This inspires us to seriously consider the system size while designing real materials. 

Having seen the plethora of dynamical states in the numerical simulations of small-sized SQUID systems, we now aim for their analytical investigation. In that respect, the particular set of equations governing the dynamics of the model allows one to explicitly construct some exact relations for small chimera and synchronous states. As an example, let us first consider the case when the fluxes in SQUIDs 1 and 2 are phase-locked but the flux through third one does not synchronize with them. This is a genuine symmetry-broken solution since Eq. (1) is symmetric under exchange of the first and the third SQUID. Writing Eq. (1) explicitly with $\phi_{1}=\phi_{2}=\phi_{0}(t)\neq \phi_{3}$,
\begin{eqnarray}
 \ddot{\phi_{1}} +\gamma\dot{\phi_{1}}+\beta sin(2\pi \phi_{1})=(\Lambda ^{-1}_{11}+ \Lambda ^{-1}_{12}+\Lambda ^{-1}_{13})
 \phi_{e}-(\Lambda ^{-1}_{11}+\Lambda ^{-1}_{12})\phi_{0}-\Lambda ^{-1}_{13}\phi_{3}\\
 \ddot{\phi_{2}} +\gamma\dot{\phi_{2}}+\beta sin(2\pi \phi_{2})=(\Lambda ^{-1}_{21}+ \Lambda ^{-1}_{22}+\Lambda ^{-1}_{23})
 \phi_{e}-(\Lambda ^{-1}_{21}+\Lambda ^{-1}_{22})\phi_{0}-\Lambda ^{-1}_{23}\phi_{3}\\
 \ddot{\phi_{3}} +\gamma\dot{\phi_{3}}+\beta sin(2\pi \phi_{3})=(\Lambda ^{-1}_{31}+ \Lambda ^{-1}_{32}+\Lambda ^{-1}_{33})
 \phi_{e}-(\Lambda ^{-1}_{31}+\Lambda ^{-1}_{32})\phi_{0}-\Lambda ^{-1}_{33}\phi_{3}.
\end{eqnarray}  
Now, for $\phi_{1}(t)=\phi_{2}(t)$, the right side of Eqs. (8) and (9) must be identical. This leads to the following algebraic relation of the flux of the synchronized pair with that of the third one
\begin{equation}
 \phi_{3}=\frac{(\Lambda ^{-1}_{11}+\Lambda ^{-1}_{13}-\Lambda ^{-1}_{22}-\Lambda ^{-1}_{23})\phi_{e}+(\Lambda ^{-1}_{22}-\Lambda ^{-1}_{11})\phi_{0}}{\Lambda ^{-1}_{13}-\Lambda ^{-1}_{23}}
\end{equation}
where we have the symmetry of the matrix $\Lambda ^{-1}$ to simplify the expression. Using the explicit form of the coupling matrix $\Lambda$ for $N=3$, we rewrite Eq. (11) in a more compact form as
\begin{equation}
 \phi_{3}=\frac{1}{8(1-\lambda_{0})}[(8-\lambda_{0})\phi_{e}+9\lambda_{0}\phi_{0}].
\end{equation} 
As noted earlier, the particular form of the coupling matrix $\lambda$ makes Eqns. (8) and (10) symmetric under the exchange of fluxes in SQUIDs 1 and 3. So, the corresponding relation for the chimera where SQUIDs 2 and 3 are synchronized is identical to Eq. (12) and reads
\begin{equation}
 \phi_{1}=\frac{1}{8(1-\lambda_{0})}[(8-\lambda_{0})\phi_{e}+9\lambda_{0}\phi_{2}].
\end{equation} 
A similar straightforward analysis fails for the third type of chimera state, but that type does exist, as seen from direct simulations (Fig. 5). However, unlike the first two types of chimeras just described, this solution preserves the symmetry of Eq. (8)-(10). Nonetheless, Eqs. (12)-(13) show that these exact results for the simplest chimera state contain all the elements of the Ashwin-Burylko criterion described above and thus theoretically confirms the possibility of small chimeras in this model. A similar analysis can be continued to systems of four and more SQUIDs in future, where there will be more diverse types of chimeras as found in Ref \cite{and2}.  

 The above analysis for chimeras can be further continued to deduce the condition for complete synchronization. For this, we set $\phi_{1}=\phi_{2}=\phi_{3}=\phi_{0}$ in Eqs. (8)-(10). This makes the left sides of Eqs. (8)-(10) identical. However, for an arbitrary value of the coupling constant $\lambda_{0}$ the right sides of the equations can be equal if and only if each of them are identically zero; i.e, $\phi_{0}=\phi_{e}$ and all the three fluxes follow the common driving flux coherently and thus gets synchronized to each other in the process. Setting $\phi_{1}=\phi_{2}=\phi_{3}=\phi_{e}$ in the right side of Eqs. (8) then yields the self-consistency relation of the driving flux 
 \begin{equation}
  \ddot{\phi_{e}} +\gamma\dot{\phi_{e}}+\beta sin(2\pi \phi_{e})=0
 \end{equation}
 as a necessary condition for complete synchronization. For our typical driving flux profile $\phi_{e}(t)=\phi_{ac}cos\Omega t$ this yields
 \begin{equation}
 2\pi\beta =\Omega ^{2},
 \end{equation}
 where we have expanded the sine in Eq. (14) upto linear order for small $\phi_{ac}$ and used the fact that $\gamma <<\Omega $ so that $tan^{-1}\frac{\Omega }{\gamma }\approx \frac{\pi }{2}$. It is easy to check that the two sides of Eq. (15) are of the same order for most typical parameter values that experimentalists work with. Hence ideal or approximate synchronization is quite common in the system and they coexist with chimera states in almost all of the parameter regime studied here. This inspires us to explore the basins of attraction for different chimera and synchronized states and this is done below. 
 
The above analytic results fail to predict the stability of the various states. To check their stability, we take resort to plotting the numerical bifurcation diagram (Fig. 6) corresponding to flux oscillation amplitudes. This is done at the equilibrium regime when the amplitudes of all the flux oscillations become steady over time. As we see from the figure, there is only one stable amplitude possible for the flux oscillations in the SQUIDs when the hysteresis parameter $\beta $ has sufficiently large values. This situation corresponds to completely synchronized states with all the three fluxes oscillating with very small amplitudes. As $\beta$ is decreased from this value, the system undergoes a bifurcation which generates an additional branch of larger possible amplitudes for the fluxes. With lowering of $\beta $ both the branches further bifurcate and thus the system becomes multistable with respect to the possible oscillation amplitudes. This continues until as $\beta$ approaches zero, when the lower branch disappears and only one solution branch remains. This multistability opens up the opportunity for the amplitudes of the three fluxes to stay on different branches and thus yield a chimera state. For larger systems, such multistability for intermediate values of the hysteresis parameter can result in the formation of discontinuities in the flux profiles as indicated in the phase diagrams of Fig. 3. Thus the bifurcation diagram showing the mechanism for generation of small chimeras may actually be valid for larger system sizes as well. On the other hand, this inspires one to look for the special features of small chimeras, which we discuss below, in larger systems as well.

 \vskip 0.5 cm
\noindent {\scriptsize {\bf B SIGNATURES OF TRANSIENT CHAOS}}
\vskip 0.5 cm
The synchronized and chimera states described in the last subsection act as attractors of the system, since they represent the asymptotic steady dynamics. As we have found, they are nonchaotic for a wide regime of parameter values, like the ones plotted in Fig. 5. However, an investigation to their full time series (Fig. 7) reveals that initially the dynamics is apparently irregular with randomly and widely fluctuating amplitudes of oscillation of the fluxes. This leads us to suspect the involvement of transient chaos in the emergence of chimera and synchronous states for the system. To confirm this, we carry out a number of studies. The simplest signature of transient chaos is the statistics of escape times \cite{tel1,tel2} and dependence of the same on initial conditions. However, to obtain the escape time, one needs to divide the total time evolution of the system into two distinct parts corresponding to initial transient dynamics and asymptotic dynamics near the attractors. In principle, these two parts can be distinguished by finite time Lyapunov spectra which typically shows different characteristics for the two different regimes \cite{tel1}. However, the current system allows for a more straightforward and numerically efficient way for this distinction. This is based on the observation that (Fig. 7) the amplitudes of flux oscillations are stable for sufficiently long time only when the system is close to the attractor and show pronounced fluctuations during the transient period. As a quantitative measure, we have detected the standard deviation of a sufficiently large (here 20) number of consecutive amplitudes and detected escape only when this is lower than a small threshold value (here 0.005). The escape time thus found is plotted against varied initial conditions. The resulting plot (Fig. 8a) indicates the irregular dependence of the escape time on initial conditions, as is the usual case \cite{tel1} for transiently chaotic systems. In particular, the discrete points on the plot with long escape times are the set of points which belong to the immediate neighborhood of the chaotic set responsible for the transient chaos. We also find the distribution of escape times for initial conditions over a chosen region in phase space (see Appendix A for details of the region). As another demonstration of transient chaos, for sufficiently long transients, the asymptotic part of the distribution is shown to be well-fitted by an exponential curve (Fig. 8b) yielding the time scale for the escape to the attractors. The time scale is dependent on the system parameters, in particular, on the damping parameter $\gamma $ which directly controls the rate of relaxation of the system. This relation, as is evident from Fig. 8c, can be fitted to a power law, with an exponent of approximately 0.92. This increase of the time scale on lowering the damping parameter can also be used to explain our earlier observation that for sufficiently small values of $\gamma $, the system fails to reach any collective state in reasonable time. This is yet another example of how the insights from the dynamics of small metamaterial systems can be successfully applied to larger system sizes as well.

\vskip 0.5 cm
\noindent {\scriptsize {\bf C BASIN OF ATTRACTION FOR A SYSTEM OF THREE SQUIDS}}
\vskip 0.5 cm
In the previous sections, we mentioned that the small chimera states and synchronized states are possible attractors at certain parameter values. However, the system governed by Eq. (1) is well-known to show pronounced multistability \cite{laz1a,laz1b,laz1c} which we explained with the bifurcation diagram of Fig. 6. For larger systems this can correspond to different initial conditions leading to a huge possibility of different chimera states, which are distinguished by the positions of the asynchronous and synchronous oscillations in the lattice and possibly non-chimera (e.g. synchronous or cluster) states. But at the level of only three SQUIDs, as explained in the earlier subsection, the only five possibilities are complete synchronization, three types of chimeras and the asynchronous states. In this section, we show that in this small system, multistability plays a dramatic role in determining the final collective state. To demonstrate this, in Fig. 9a and 9b, we plot the basins of attraction of the system projected in the planes of initial conditions spanned by $\phi_{1}(0)-\phi_{2}(0)$ (with $\phi_{3}(0)$ fixed at 0) and $\phi_{1}(0)-\phi_{3}(0)$ (with $\phi_{2}(0)$ fixed at 0) both with $\dot\phi_{i}(0)=0$ for $i=1,2,3$ (the actual phase space of the second-order system Eq. (8)-(10) with $N=3$ is six-dimensional). To plot the basins, we detected a chimera if $|R|$ calculated for a state and averaged between times $4800-5000$ was between $0.4-0.8$. A greater time-averaged $|R|$ was marked as synchronous, while a lesser was marked as some other state. In this respect, we note that, unlike the case for bigger systems, $|R|$ can now successfully identify a small chimera since it yields an intermediate value only if any two of the fluxes are synchronous and that always corresponds to a chimera state. 

The basins of attraction are seen to be composed of both regular and apparently random components and have interesting structures. The regular parts would correspond to predictability of final states, while for the other parts, a minute change in the initial conditions may lead to a very different final state. The relatively regular parts in the plots occur in the vicinity of the origin and along the lines on which two initial fluxes are equal, except for the line $\phi _{1}(0)=\phi _{3}(0)$ (Fig. 9a and 9b). In the former case, all three fluxes start with magnitudes very close to each other, so that they ultimately synchronize. In the latter cases, any two of the fluxes start with values close to each other, such that the initial configurations themselves resemble a chimera state. However, as is seen from the figures, this may or may not ultimately lead to a chimera state since they can be totally synchronized also. Interestingly, the initial conditions responsible for synchronized and chimera states in such cases are seen to be arranged in a regular manner with well-defined periodicity which also slowly varies with the distance from the origin (Fig. 9a and 9b). This regularity vanishes sufficiently away from the origin, which indicates that, for initial fluxes of significantly larger values, the system is not predictable at all, even if two of the initial fluxes are equal. This extreme
sensitivity to initial conditions is a well-known property of chaotic systems \cite{chaos1,agu1}, where these are responsible for fractal basin boundaries. In those cases, the trajectories of the system behave chaotically near the vicinity of a nonattracting chaotic set in the phase space and, finally escape it via different routes and consequently terminate on different attractors. So, this behavior can be yet another manifestation of the involvement of initial transient chaos in the system's final relaxation to an attractor. This idea can be carried over to systems of larger sizes with a richer diversity of attractors and the involvement of transient chaos to the generation of chimera states in other systems can be studied. In fact, it is well-known \cite{kur1,omel1} that, for spatially extended systems, the position of the synchronous portion in the lattice is strongly dependent on the initial condition. This might be an indication that transiently chaotic dynamics is involved in shaping a chimera's spatial profile in any general system.  
 
 The study of fractal basins in dissipative systems is a relatively less explored topic with generic predictions being done only recently \cite{xiao1,adil1,tel2}. In this respect, a particularly interesting type of fractal basin is the so called \textit{Wada basin} where basins corresponding to three or more attractors meet at the same boundary \cite{agu1,xiao1}. This interesting topological structure has been discovered in certain chaotic systems \cite{agu1,xiao1,ble1} and conditions for its appearance are being developed \cite{agu1,xiao1}. In our system, we check the fractality and Wada property of our basins through direct numerical simulation. To do this, we have adapted the method in Ref \cite{xiao1}, while a different method is discussed in Ref. \cite{alv1}. To confirm the fractal and Wada properties, we have selected a portion of the basin boundary and zoomed in parts of it in successive steps. As is seen from the typical example in Fig. 9c, as we keep on increasing the resolution, an apparent boundary between two basins actually shows portions of other basins inside it at higher resolution and it continues until very small scales. To ensure the numerical precision, we choose a step size of $dt=0.002$ in doing this. However, continuing this to arbitrary fine length scales would really be an interesting theoretical study, but not of interest from experimental considerations where one has limited control over initial flux adjustments. Nonetheless, the studies in this section establish the system as an experimentally realizable example of dissipative system showing sensitive dependence on initial conditions and fractal Wada basins, thereby encouraging the experimental study of dissipative chaos.

\vskip 1cm
\noindent {\bf IV. CONCLUSIONS}
\vskip 0.5 cm

In the preceding sections, we described how the small chimera states emerge out of transient chaos in a model of nonlinear metamaterial. These studies, in one hand, encourage to study the mechanism behind emergence of chimeras in other models, and in the other, establish the superconducting metamaterial as a unique system to experiment with nonlinear dynamics. We now conclude by mentioning some limitations of the study and offering possible directions of future works.

 The effective model used to simulate the material has many idealizations and accommodations of realistic effects may have significant effect on collective dynamics. Firstly, our considerations here neglected any imperfections in the system coming from, e.g, critical current fluctuations of the SQUIDs \cite{gra1}, intrinsic low-frequency flux noises due to spin-diffusion \cite{lan1} or local magnetic impurities \cite{bial1}, and other $1/f$ noise sources \cite{pala1}. Furthermore, the characterizations of all the individual SQUIDs in the material may not be identical and so ideally there should be a distribution of the $\beta $,$\gamma $ and $\lambda _{0}$ parameters. The external magnetic field may not be temporally coherent and its spatial variations may not be negligible. Furthermore, as was shown in Ref. \cite{laz1b}, considerations of nonzero initial flux velocities allow more diversity and multistability in the dynamics with prominent multichimera states containing varied number of synchronous groups. We neglected this possibility throughout our studies.

Our present work complements the identification of chimera states as transiently chaotic states themselves and shows that even in the self-organization of these states, transient chaos is involved. To check if this relation is robust, one needs to study other chimera-sustaining systems and discern the characteristics of the transients leading to chimera states there. Further sophisticated studies may involve investigations of the finite-time Lyapunov spectra at various times. This would be particularly needed if the chimera states themselves are chaotic and then one may like to observe the differences in the Lyapunov spectra corresponding to the transient part, the fully developed chimera states and (if present) the state obtained after the chimera state collapses. For extended systems, the transient chaos may be spatial, temporal or spatiotemporal and thus offer much richer dynamical possibilities.  Since currently both chimeras and chaos enjoy a sound establishment as prominent contemporary research topics and both exhibit a certain level of ubiquity in natural and model systems, we hope that the newly found relation between them will open up novel directions of future work in the field of nonlinear dynamics.   

\vskip 1cm
\noindent {\bf ACKNOWLEDGMENTS}
\vskip 0.5 cm
AB acknowledges the support by the Kishore Vaigyanik Protsahan Yojanaa (KVPY) program and DS acknowledges the support by the Innovation in Science Pursuit for Inspired Research (INSPIRE) program, both from the Department of Science and Technology, Government of India. Useful and encouraging discussions with Prof. Jayanta Kumar Bhattacharjee and Prof. Rajarshi Roy are gratefully acknowledged. The authors are thankful to Prof. Tamas Tel for reading the manuscript and providing encouraging and insightful feedback.

\vskip 1cm
\noindent {\bf APPENDIX A: Phase Space for Escape Time Statistics}
\vskip 0.5 cm
Using stagger-and-step method discussed in \cite{tel1}, we found out a trajectory which spends sufficiently long time near the chaotic set. That trajectory had initial conditions: $\phi_{1}(0) =-0.46000000034207056$, $\phi_{2}(0)=-0.99999999937360939$, $\phi_{3}(0)=4.0195482530189722E-010$, $\dot\phi_{1}(0)=-1.8923665127134672E-010$, $\dot\phi_{2}(0)=1.1154667603405408E-010$, $\dot\phi_{3}(0)=5.7399544471812762E-010$ and had an escape time of about 5058 units. To plot Fig. 9a, we varied $\phi_{1}(0)$ keeping other initial conditions fixed at the mentioned values. For the histogram in Fig. 9b, the phase space region used for sampling initial conditions is a six dimensional sphere of radius $10^{-4}$ units centered at the mentioned point, and the points contained within it are chosen from a uniform distribution. To plot the escape time histogram, only the trajectories having escape times more than 1500 units are considered.

\begin{center}

--------------------------------------------------------------------------------------------------------------------------
\end{center}

\newpage
\begin{center}
 \textbf{\underline{Figures}}
\end{center}
\begin{figure}[ht]
\begin{center}
\begin{tabular}{c}
        \resizebox{10cm}{!}{\includegraphics[angle=0]{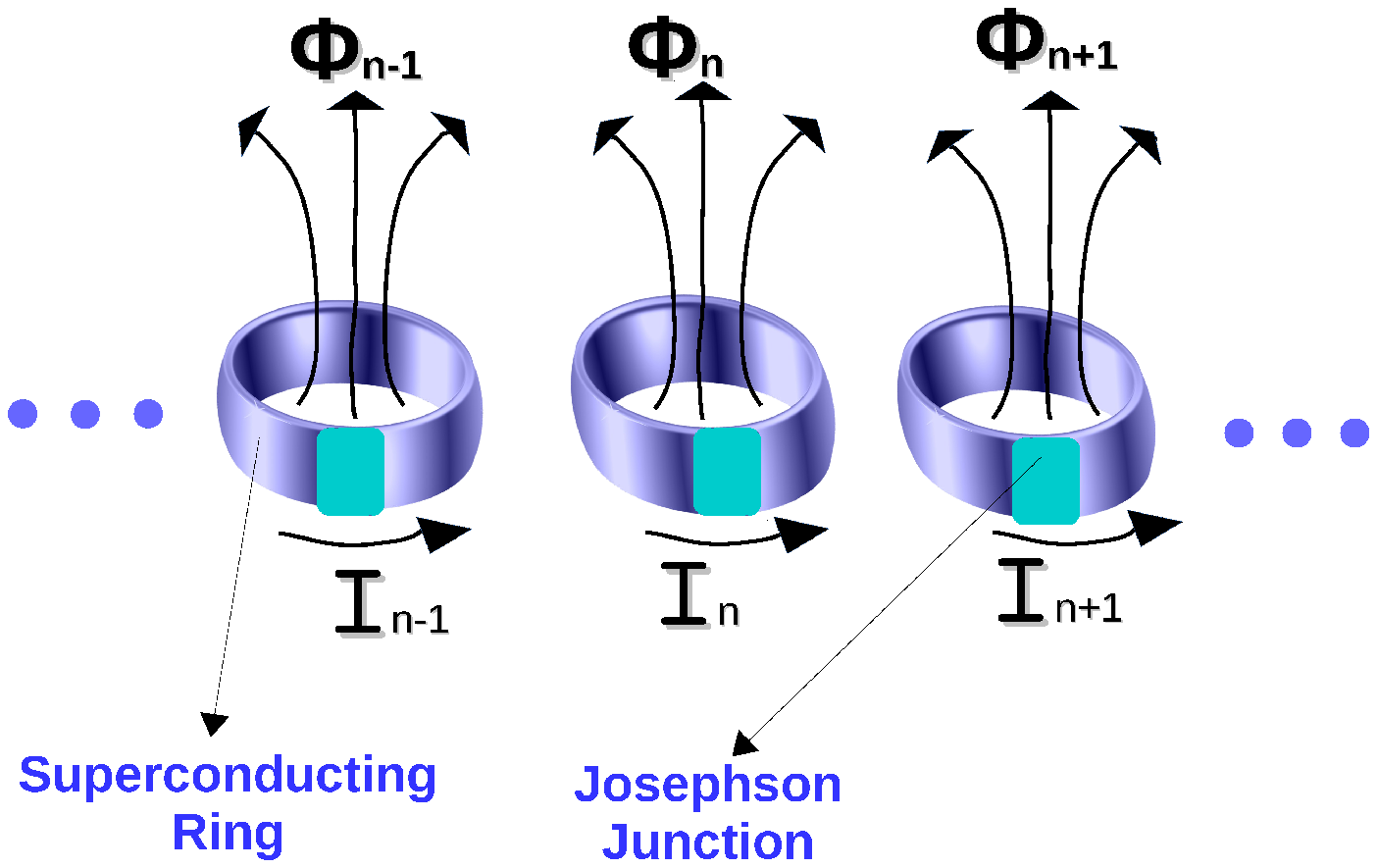}}
        
                  \end{tabular}
 \caption{{(Color online) Schematic diagram for the currents ($I_{n}$) and fluxes ($\phi _{n}$) in the SQUID metamaterial.}}
\end{center}
\end{figure}
\vskip 4 cm
\begin{figure}[h]
\begin{center}
\begin{tabular}{c}
        \textbf{Asynchronous}\\
        \textbf{(a)}\resizebox{8cm}{!}{\includegraphics[angle=-90]{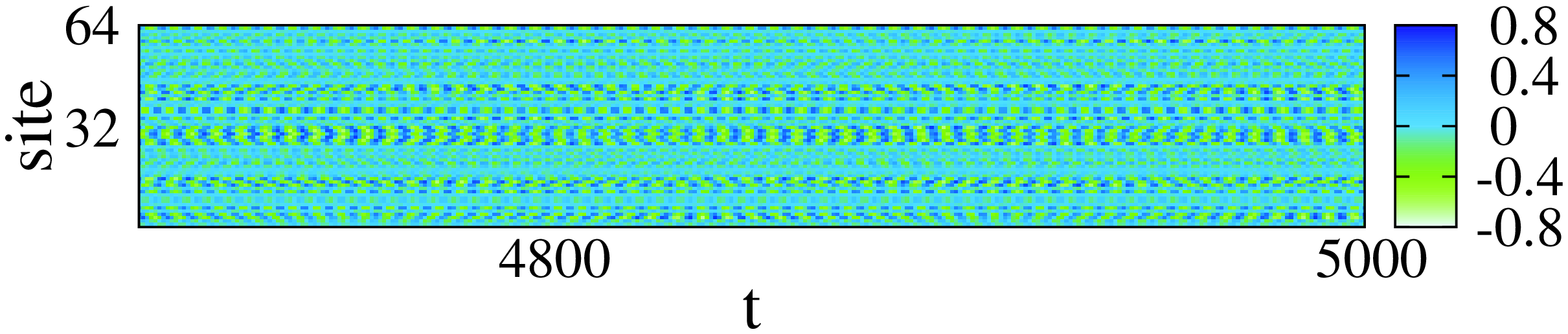}}\\
                \textbf{Chimera}\\
        \textbf{(b)}\resizebox{8cm}{!}{\includegraphics[angle=-90]{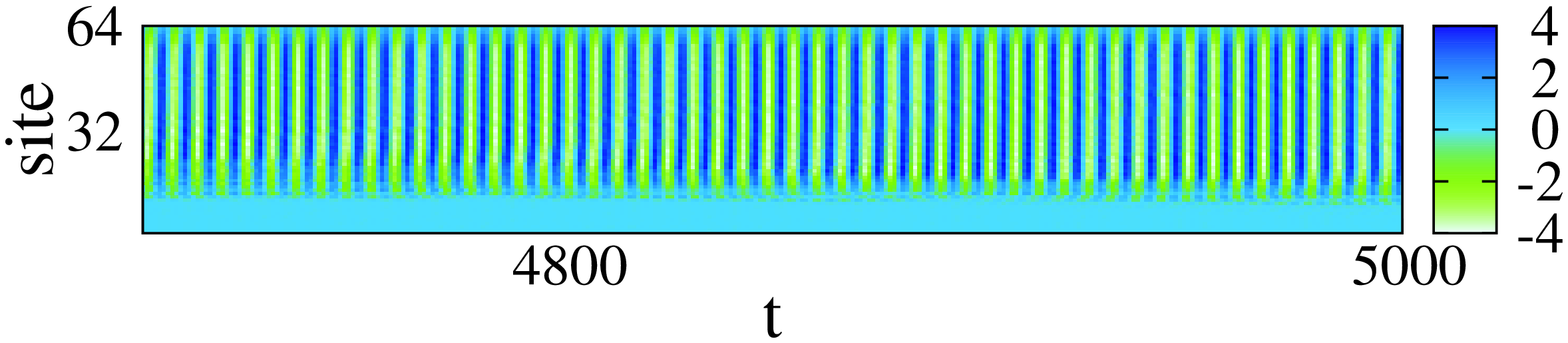}}\\
                \textbf{Multichimera}\\
        \textbf{(c)}\resizebox{8cm}{!}{\includegraphics[angle=-90]{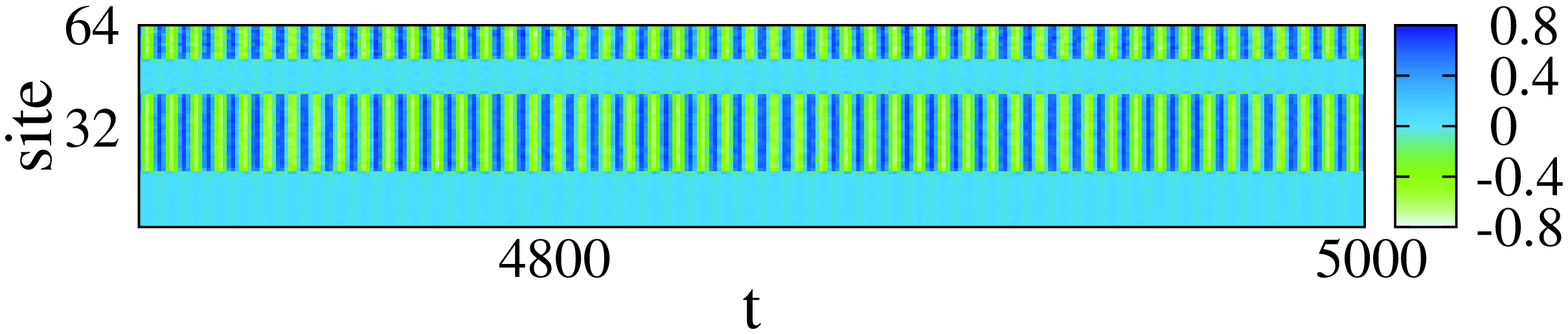}}\\
                \textbf{Cluster}\\
        \textbf{(d)}\resizebox{8cm}{!}{\includegraphics[angle=-90]{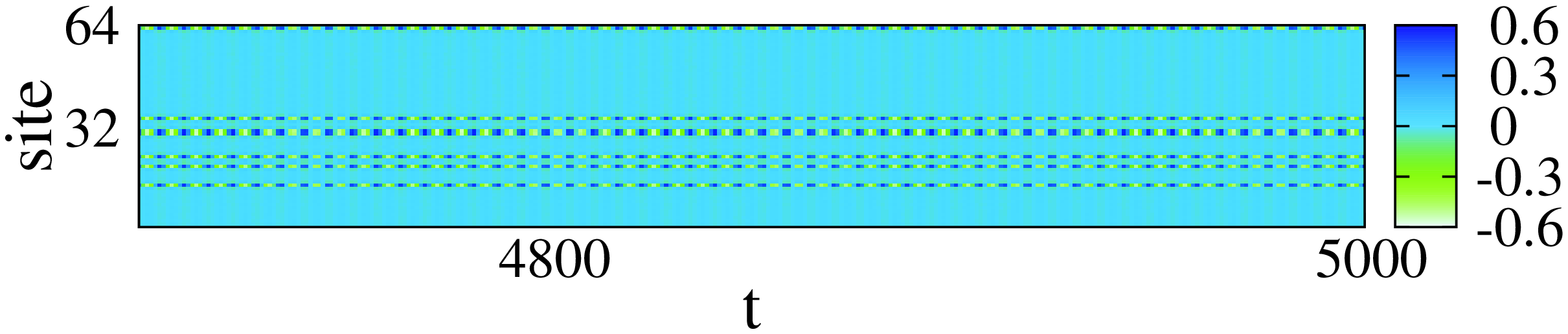}}\\
        \textbf{(e)}\resizebox{8cm}{!}{\includegraphics[angle=-90]{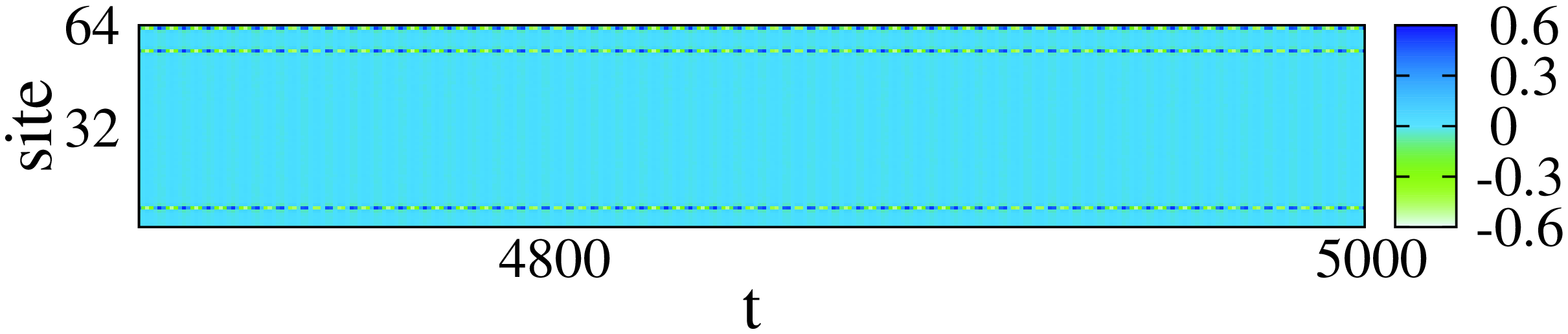}}\\
                \textbf{Synchronized}\\
        \textbf{(f)}\resizebox{8cm}{!}{\includegraphics[angle=-90]{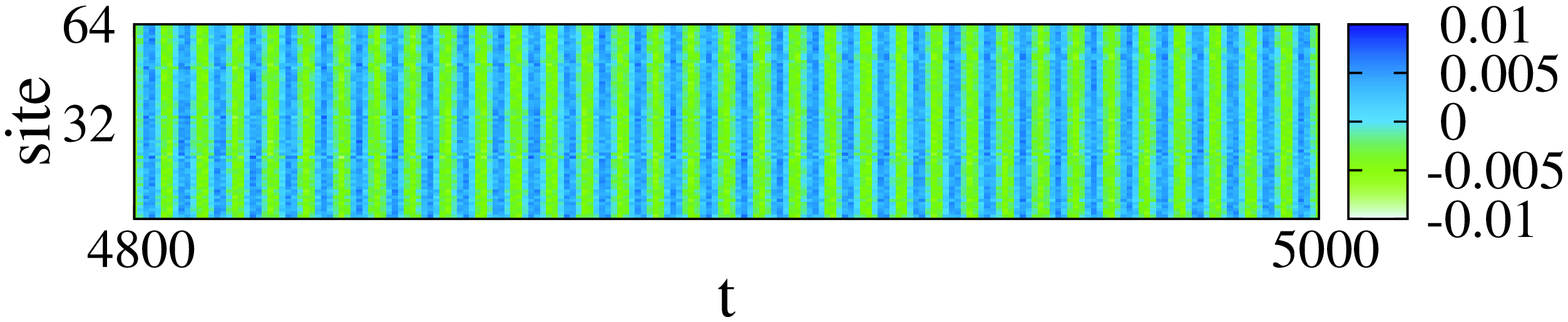}}
        
                  \end{tabular}
 \caption{{(Color online) Space-time plots for different collective modes of fluxes, all self-organized from identical random initial flux configuration, with $N=64, \phi _{R}=0.9,\phi _{ac}=0.01,\Omega =2\pi /5.9,dt=0.02$, (a) asynchronous: $\beta =0.1,\gamma =0.0001,\lambda _{0}=-0.005$; (b) chimera: $\beta =0.088,\gamma =0.002,\lambda _{0}=-0.05$; (c) multichimera: $\beta =0.09,\gamma =0.002,\lambda _{0}=-0.05$; (d) cluster: $\beta =0.125,\gamma =0.002,\lambda _{0}=-0.05$; (e) cluster: $\beta =0.139,\gamma =0.002,\lambda _{0}=-0.05$; (f) synchronous: $\beta =0.4,\gamma =0.002,\lambda _{0}=-0.05$.}}
\end{center}
\end{figure}

\begin{figure}[ht]
\begin{center}
\begin{tabular}{c}
        \textbf{(a)}\resizebox{8cm}{!}{\includegraphics[angle=-90]{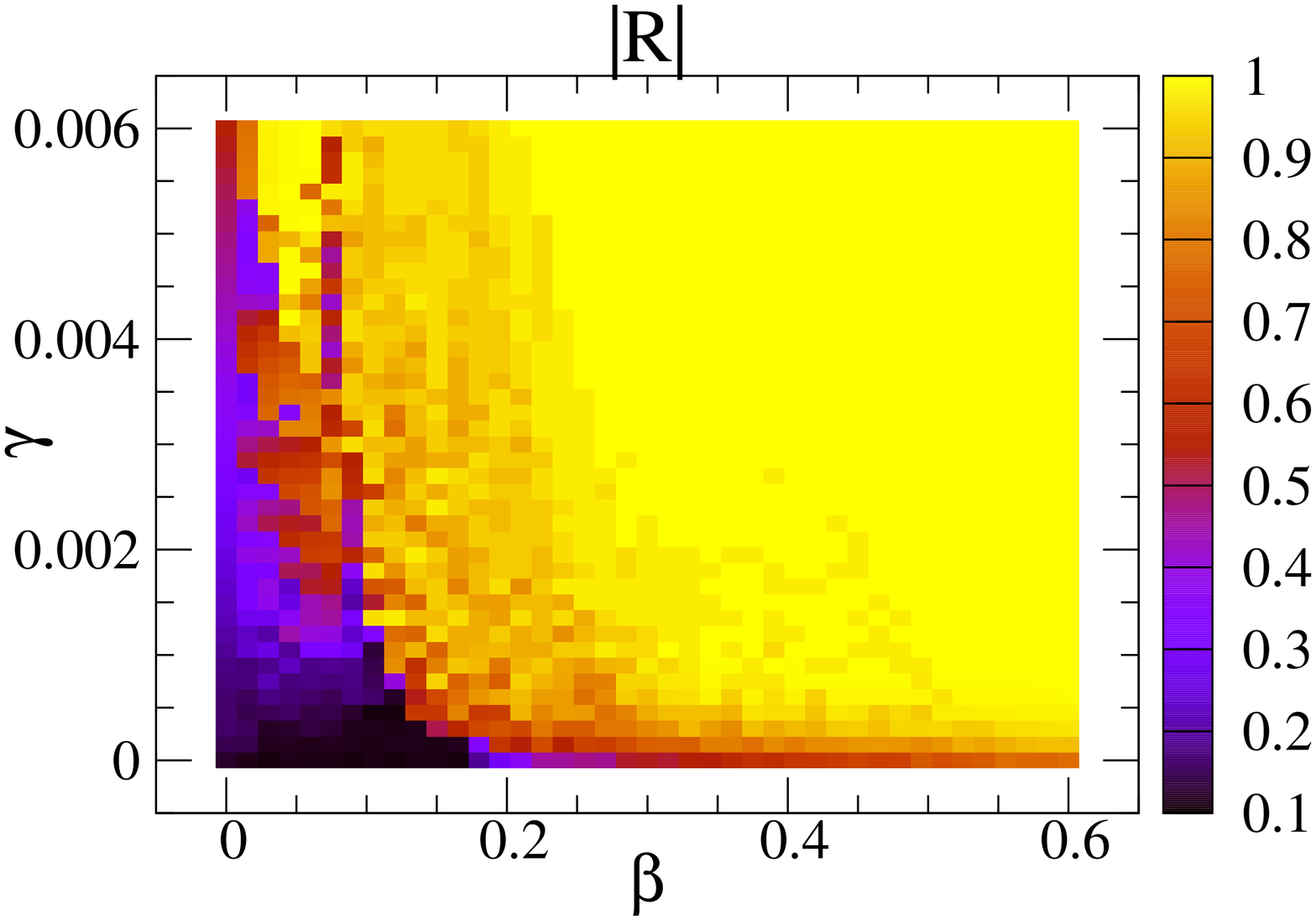}}\\
        \textbf{(b)}\resizebox{8cm}{!}{\includegraphics[angle=-90]{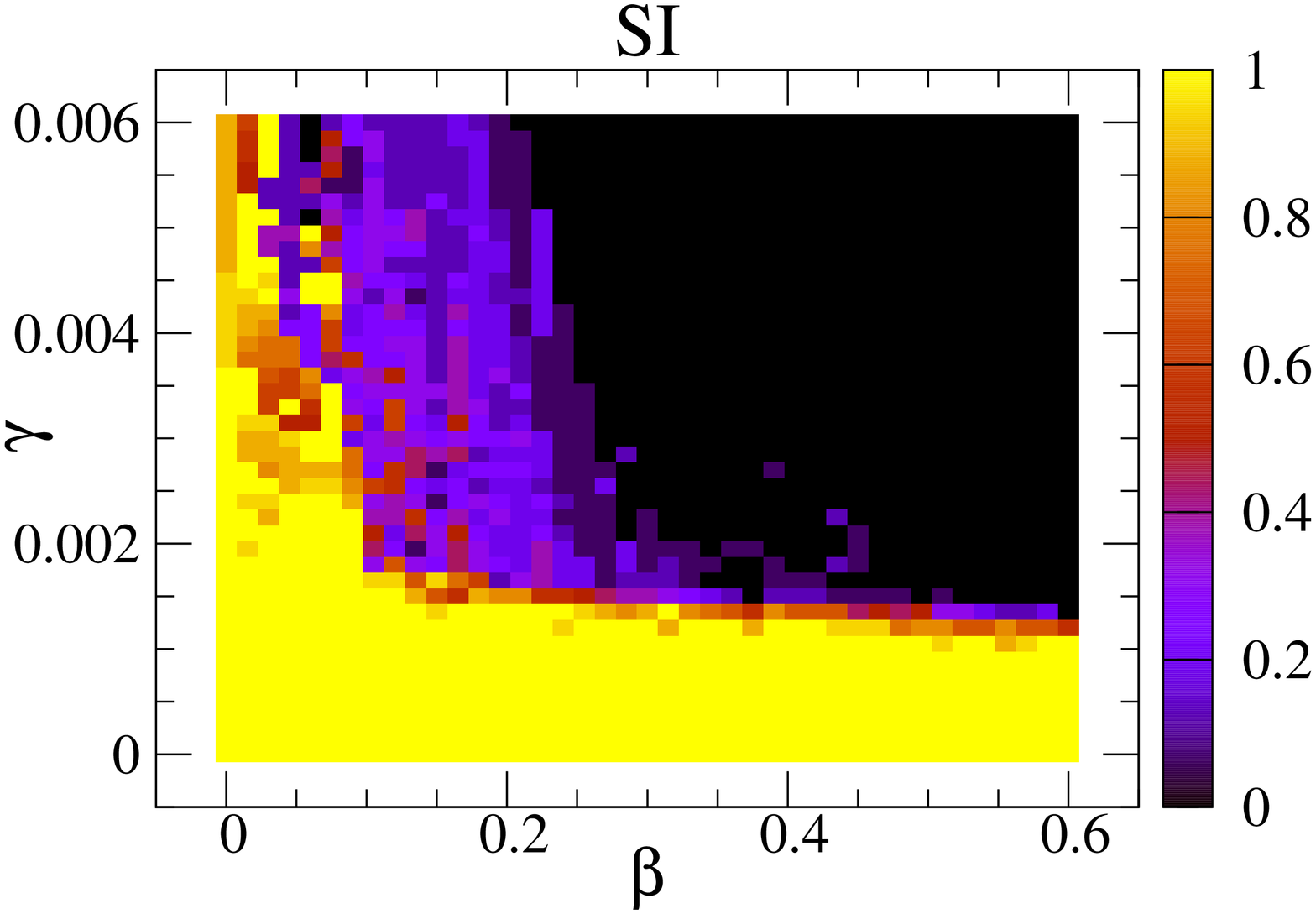}}\\
        \textbf{(c)}\resizebox{8cm}{!}{\includegraphics[angle=-90]{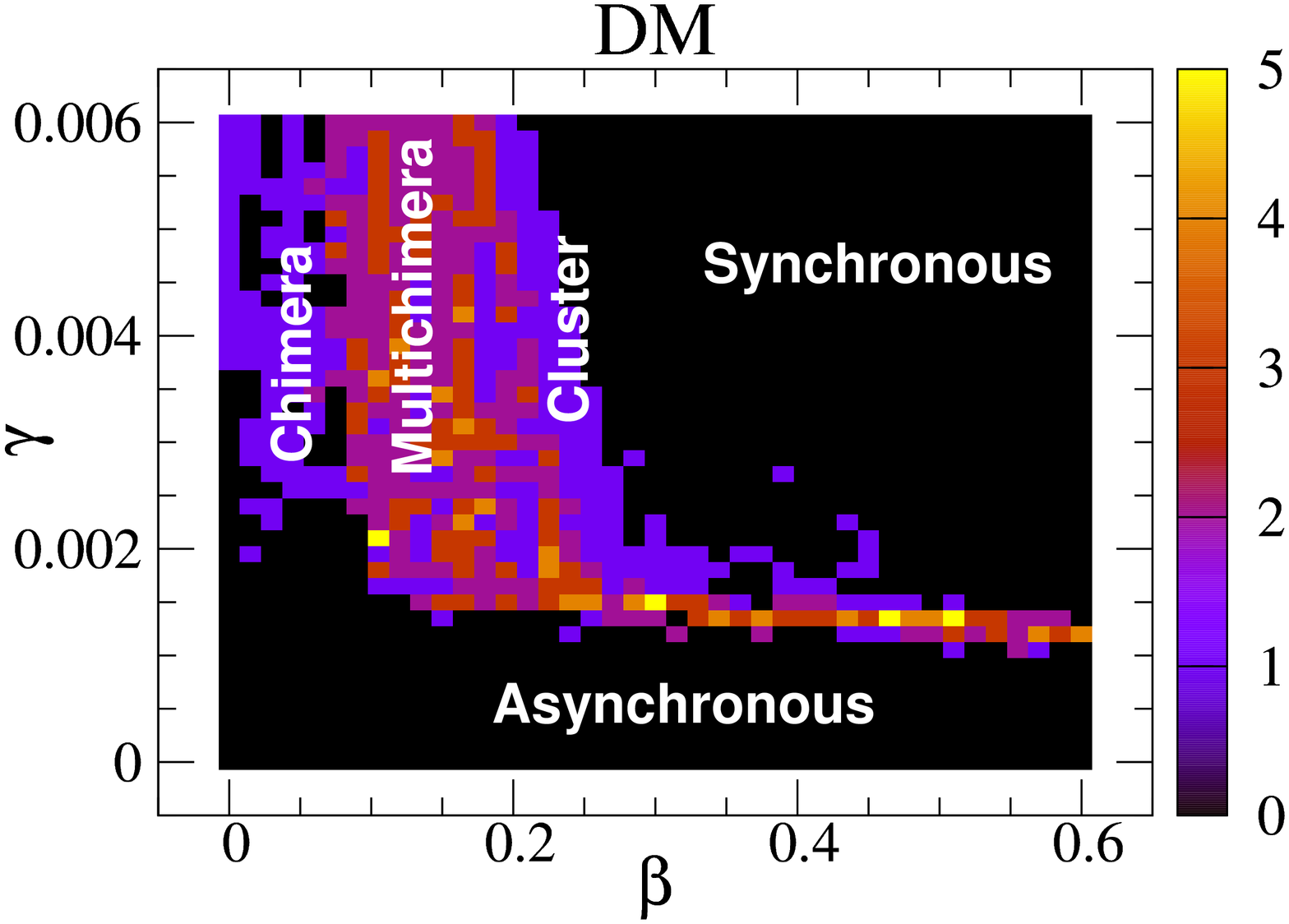}}
        
                  \end{tabular}
 \caption{{(Color online) Phase diagrams for (a) $|R|$, (b) SI and (c) DM starting from random but identical initial flux configuration, with $R$ and local standard deviations time-averaged between 4800-5000 time units, $N=64,N_{g}=16,\delta =0.005,\lambda_{0}=-0.05,\phi _{R}=0.9,\phi _{ac}=0.01,\Omega =2\pi /5.9,dt=0.02$.}}
\end{center}
\end{figure}

\begin{figure}[ht]
\begin{center}
\begin{tabular}{c}
        \textbf{(a)} \resizebox{8cm}{!}{\includegraphics[angle=-90]{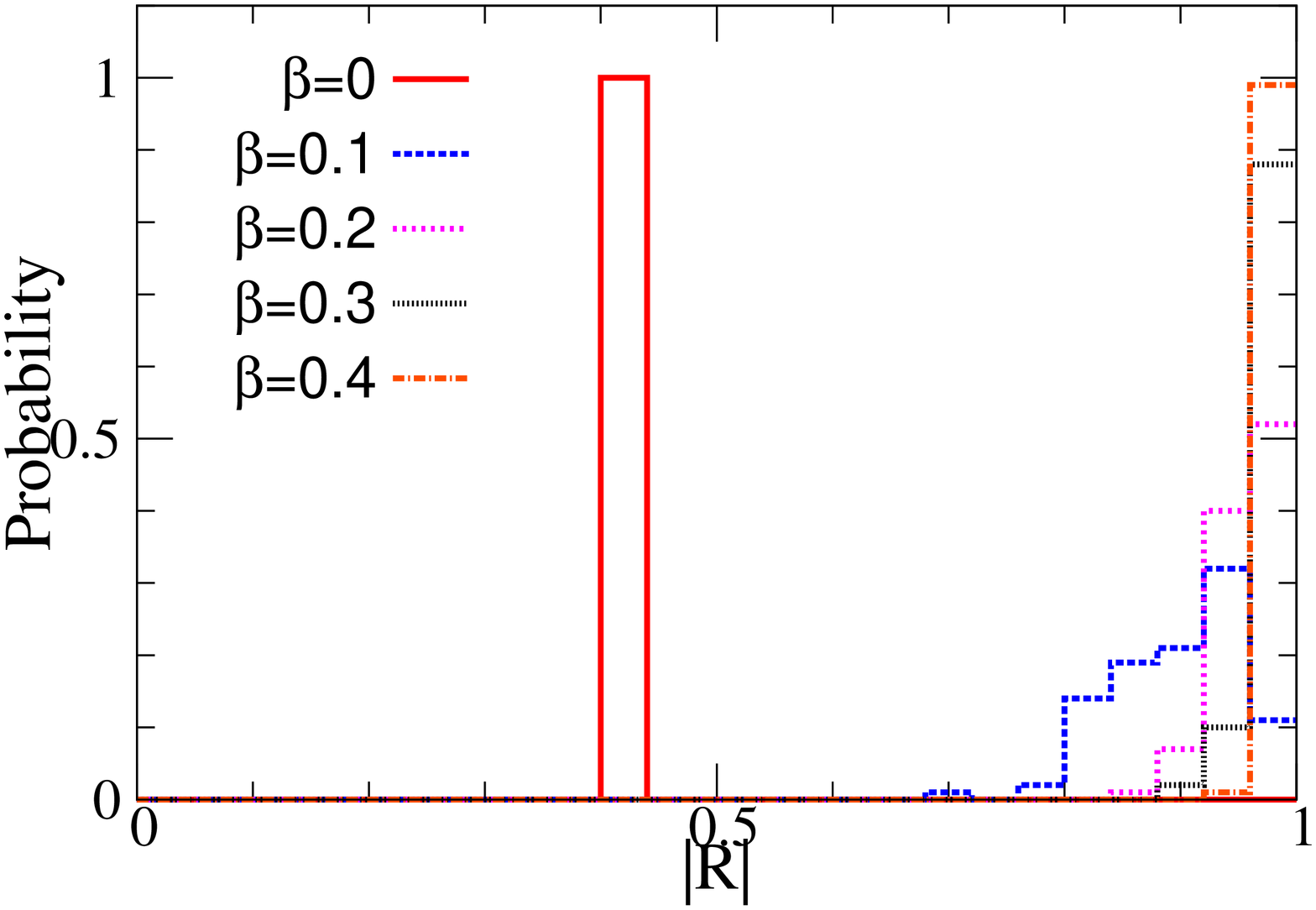}}

\end{tabular}

 \caption{{(Color Online) Histograms for the modulus of the order parameter $R$ calculated at time 5000 unit with $N=64,dt=0.02,\phi _{R}=0.9,\gamma=0.004,\lambda_{0}=-0.05$ each for an ensemble of 100 random initial flux distributions.}}
\end{center}
\end{figure}

\begin{figure}[h]
\begin{center}
\begin{tabular}{c}
        \textbf{Small Chimera}\\
        \textbf{(a)}\resizebox{8cm}{!}{\includegraphics[angle=-90]{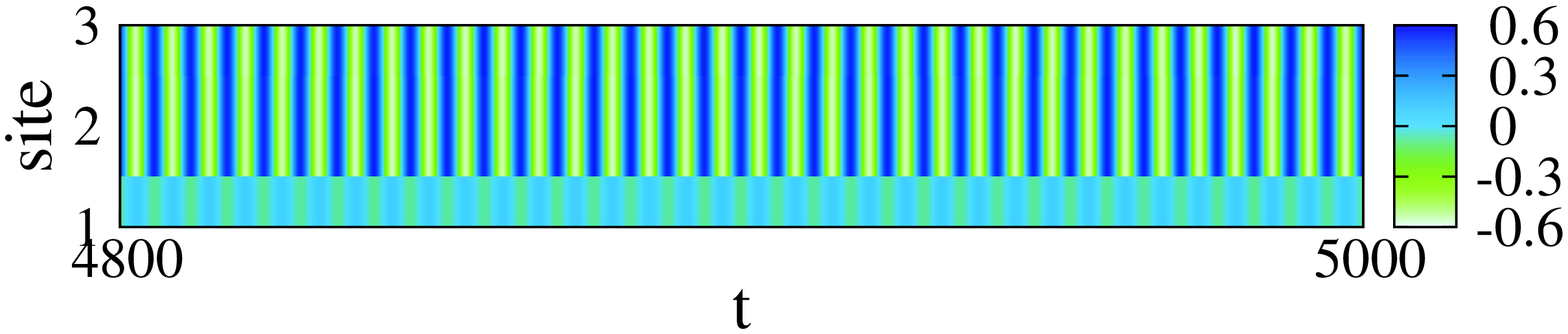}}\\
        \textbf{(b)}\resizebox{8cm}{!}{\includegraphics[angle=-90]{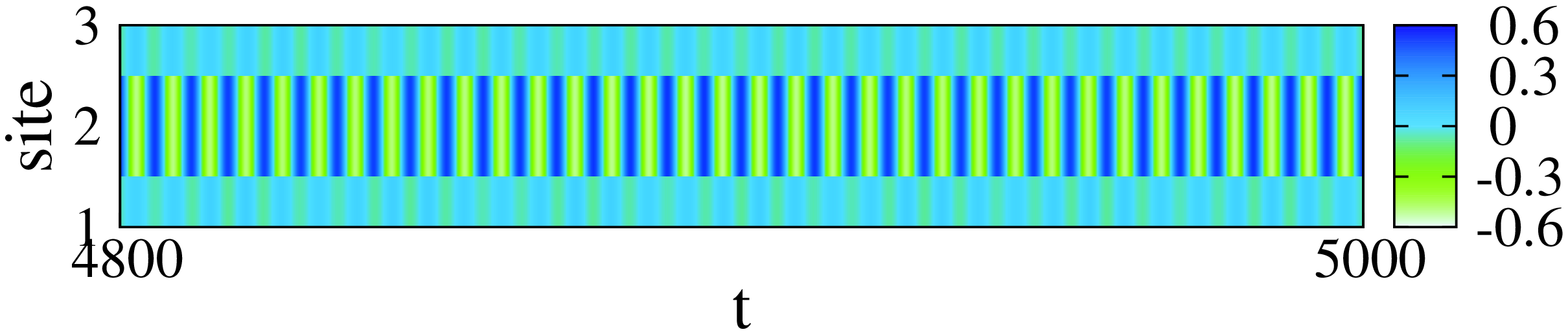}}\\
        \textbf{(c)}\resizebox{8cm}{!}{\includegraphics[angle=-90]{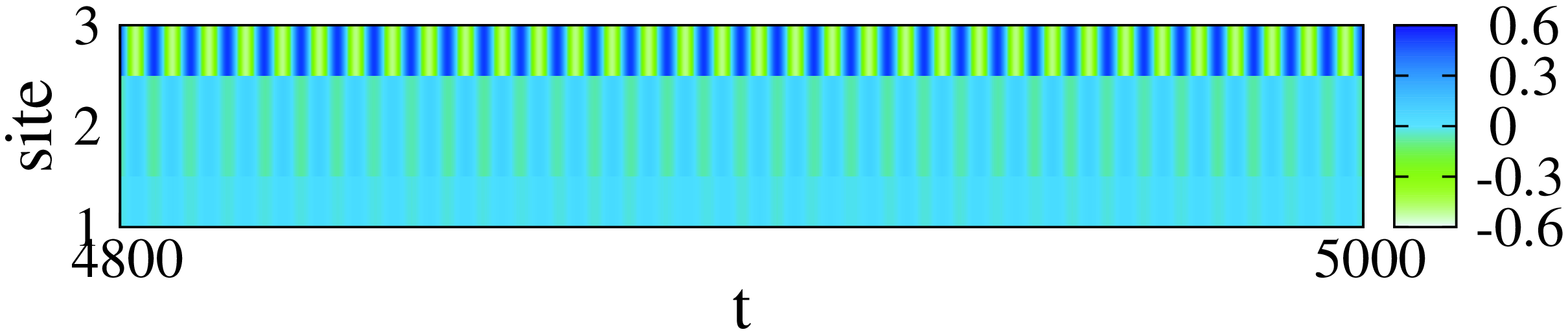}}\\
         \textbf{Synchronized}\\
        \textbf{(d)}\resizebox{8cm}{!}{\includegraphics[angle=-90]{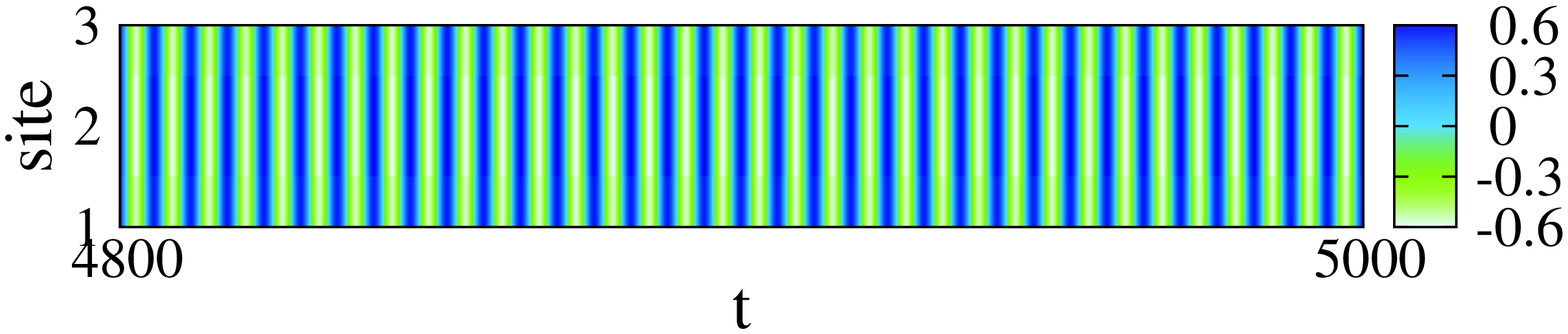}}\\
        \textbf{Asynchronous}\\
        \textbf{(e)}\resizebox{8cm}{!}{\includegraphics[angle=-90]{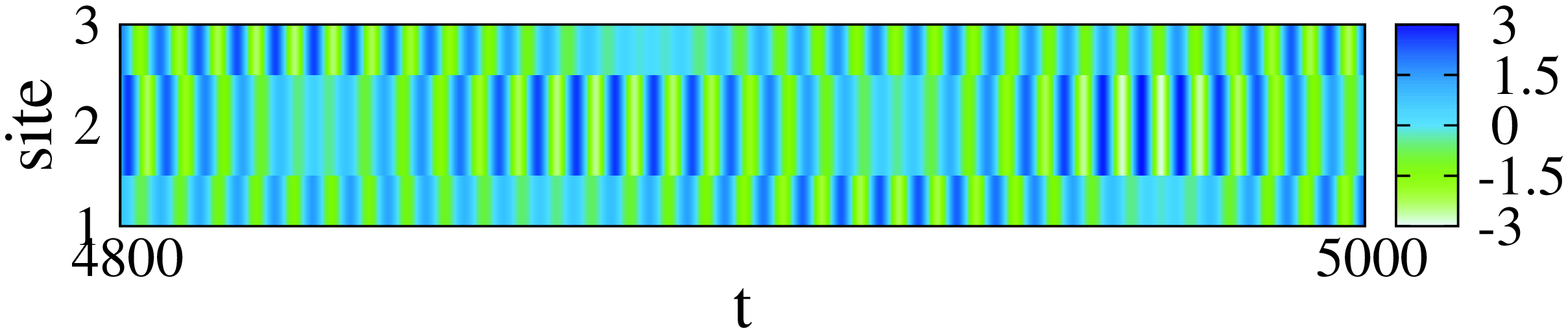}}
        
                  \end{tabular}
 \caption{{(Color online) Space-time plots for (a)-(c) small chimera, (d) synchronous and (e) asynchronous states for $N=3$. In all the plots $\beta =0.1114,\lambda _{0}=-0.05,\phi _{ac}=0.01,\Omega =2\pi /5.9,dt=0.02,\dot\phi _{n}(0)=0$ for $n=1,2,3$, $\phi_{1}(0)=-2.5$ and $\phi_{3}(0)=0$ whereas (a) $\gamma =0.002,\phi_{2}(0)=-2.5$, (b) $\gamma =0.002,\phi_{2}(0)=-2.28$, (c) $\gamma =0.002,\phi_{2}(0)=-2.38$, (d) $\gamma =0.002,\phi_{2}(0)=-2.48$ and (e)$\gamma =0,\phi_{2}(0)=-2.5$.}}
\end{center}
\end{figure}
\begin{figure}[h]
\begin{center}
\begin{tabular}{c}
        \resizebox{8.5cm}{!}{\includegraphics[angle=-90]{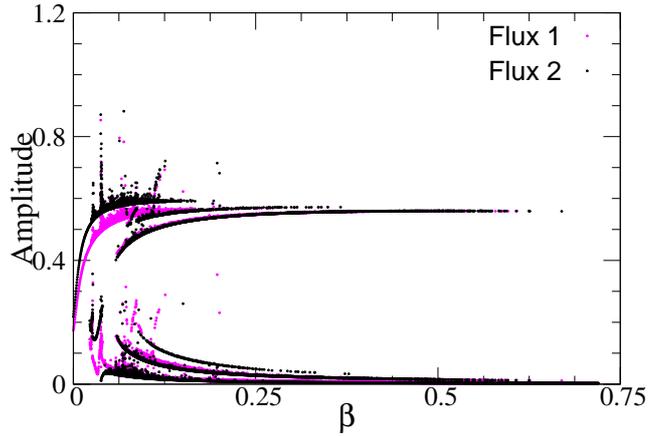}}

                  \end{tabular}
 \caption{{(Color online) Bifurcation diagrams for amplitudes of flux oscillations, between time 4800-5000 with $\dot\phi _{1}(0)=\dot\phi _{2}(0)=\dot\phi _{3}(0)=0,\gamma =0.002,\lambda _{0}=-0.05,\phi _{ac}=0.01,\Omega =2\pi /5.9, dt=0.02$. Initial flux values are taken from the range $[-0.45,0.45]$. Due to symmetry, the corresponding plot for $\phi _{3}$ is identical to that of $\phi _{1}$ and hence not shown here.}}
\end{center}
\end{figure}
\begin{figure}[h]
\begin{center}
\begin{tabular}{c}
        \textbf{(a)}\resizebox{8cm}{!}{\includegraphics[angle=0]{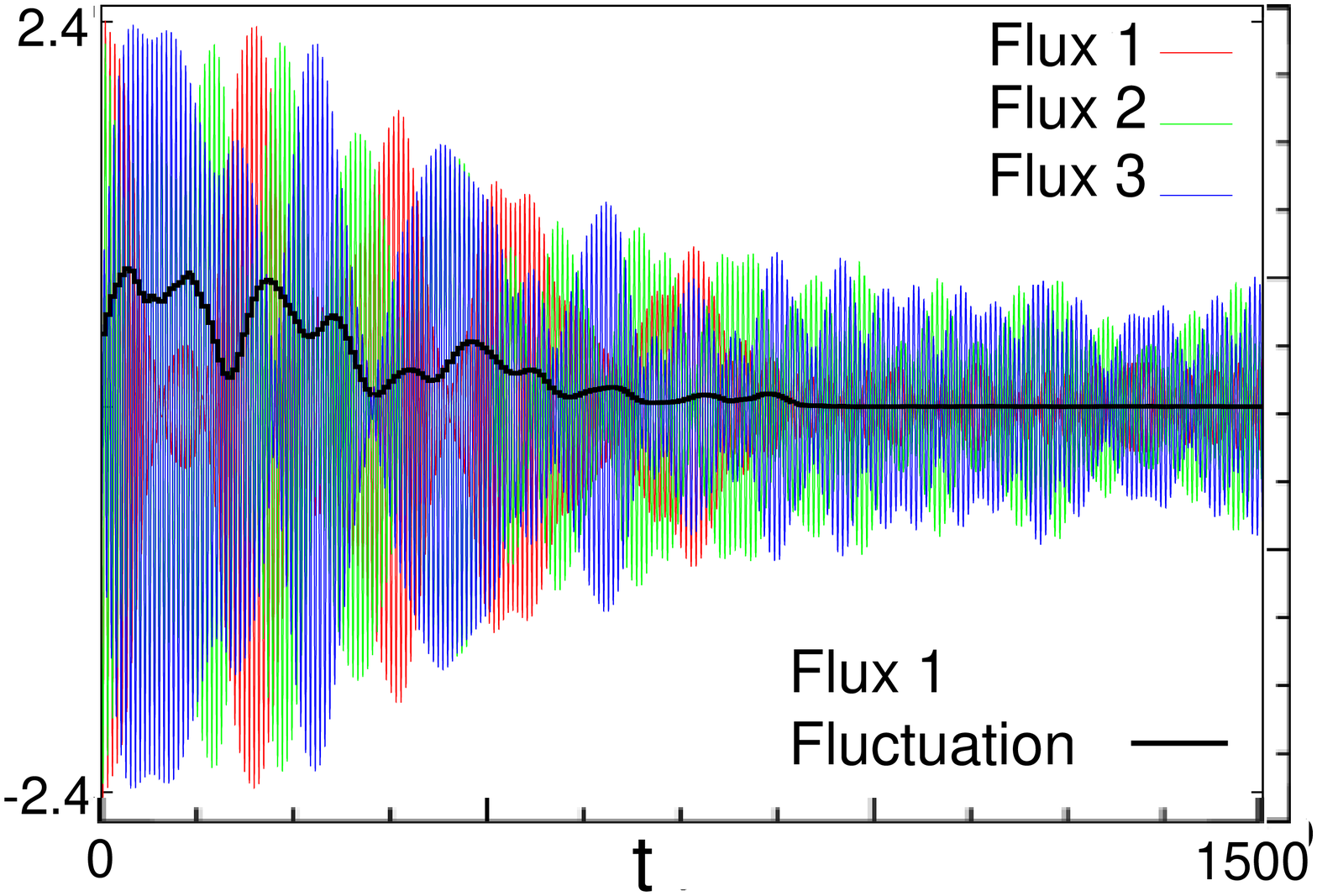}}\\
        \textbf{(b)}\resizebox{8cm}{!}{\includegraphics[angle=0]{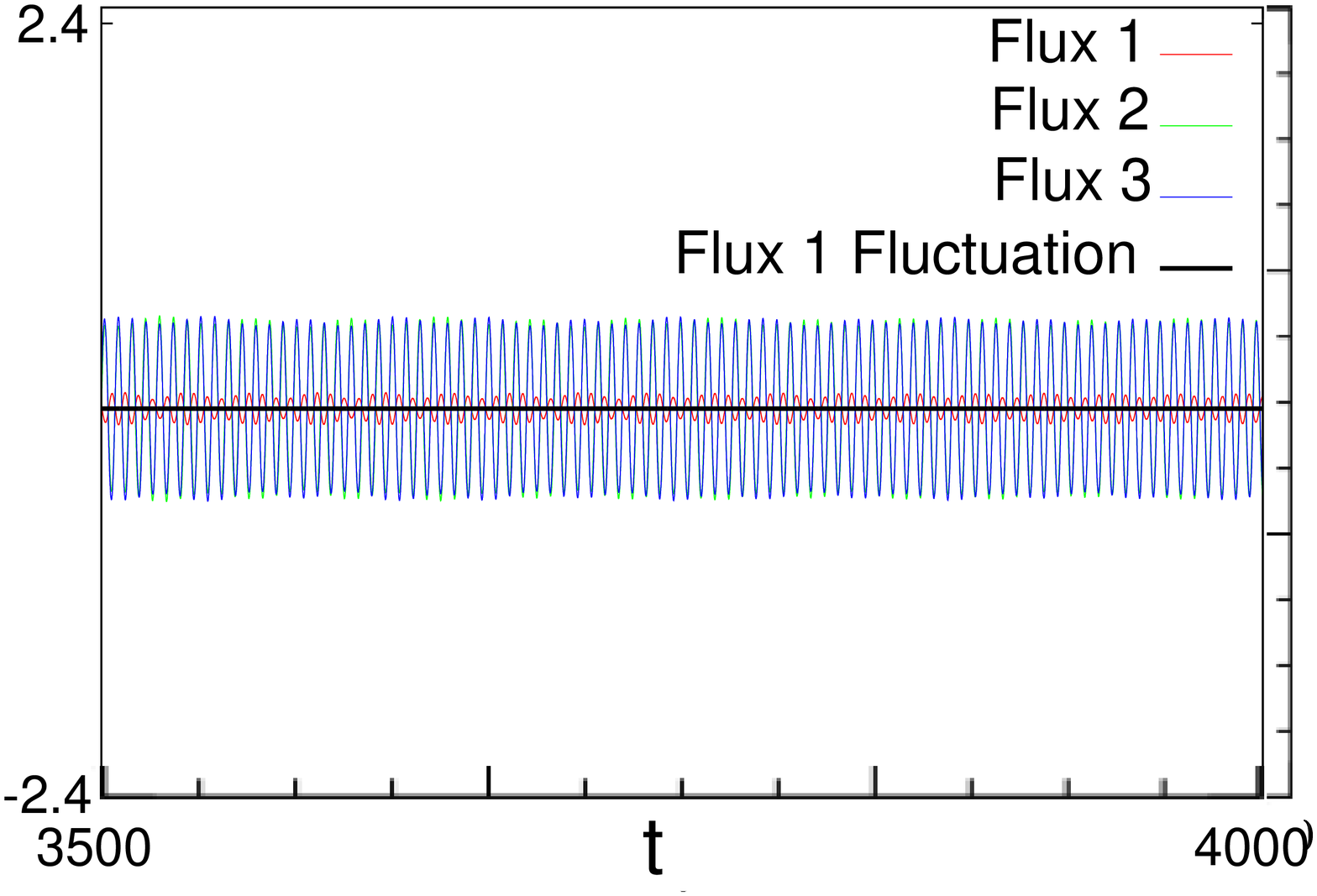}}

                  \end{tabular}
 \caption{{(Color online) (a) Transient and (b) asymptotic time series and amplitude fluctuations (standard deviation) for the evolution of a chimera state with $\phi _{1}(0)=\phi _{2}(0)=-2.5,\phi _{3}(0)=\dot\phi _{1}(0)=\dot\phi _{2}(0)=\dot\phi _{3}(0)=0,\beta =0.1114,\gamma =0.002,\lambda _{0}=-0.05,\phi _{ac}=0.01,\Omega =2\pi /5.9, dt=0.02$.}}
\end{center}
\end{figure}

\begin{figure}[ht]
\begin{center}
\begin{tabular}{c c}
        \textbf{(a)}\resizebox{8cm}{!}{\includegraphics[angle=-90]{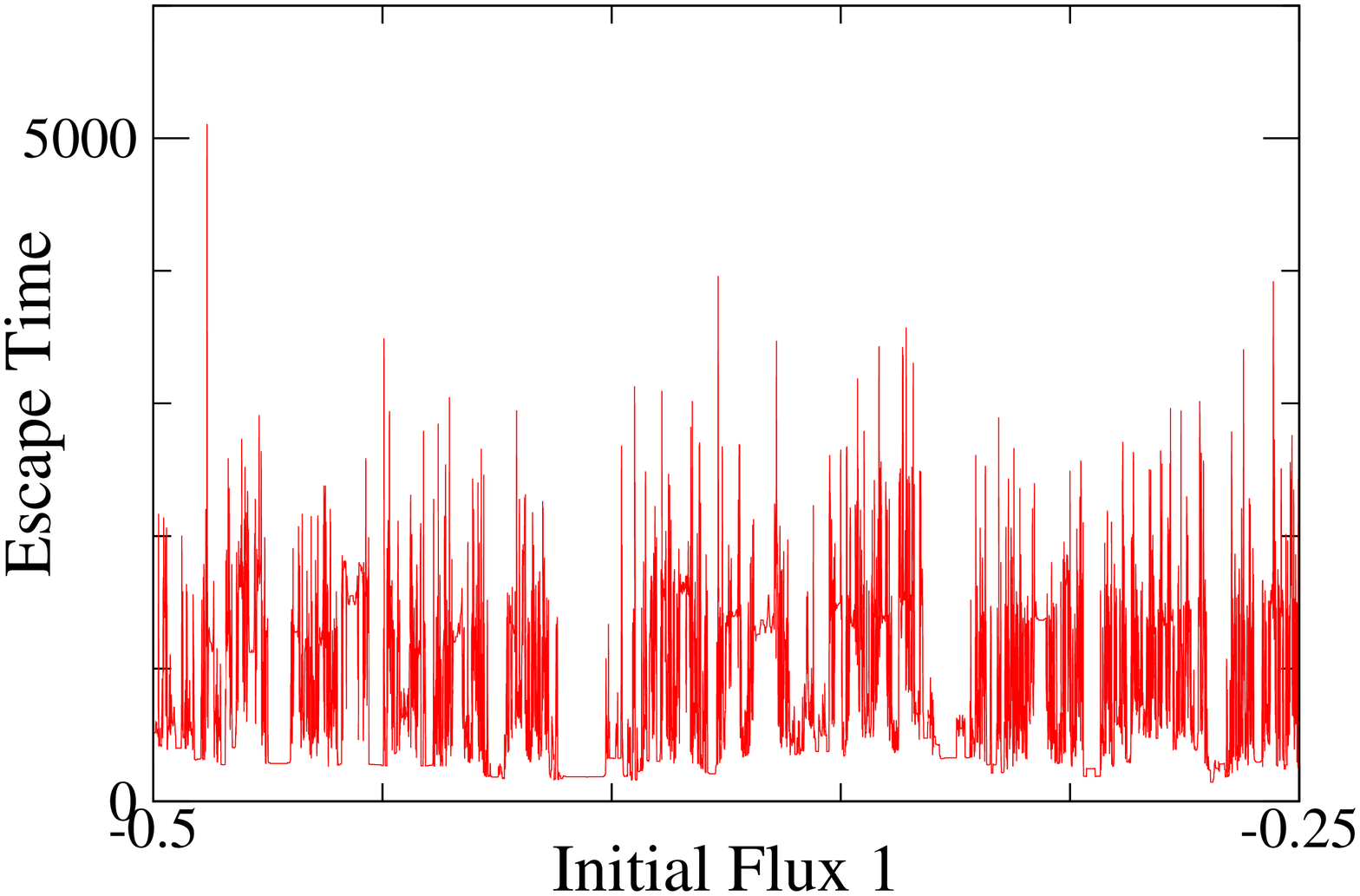}}\\
        \textbf{(b)}\resizebox{8cm}{!}{\includegraphics[angle=90]{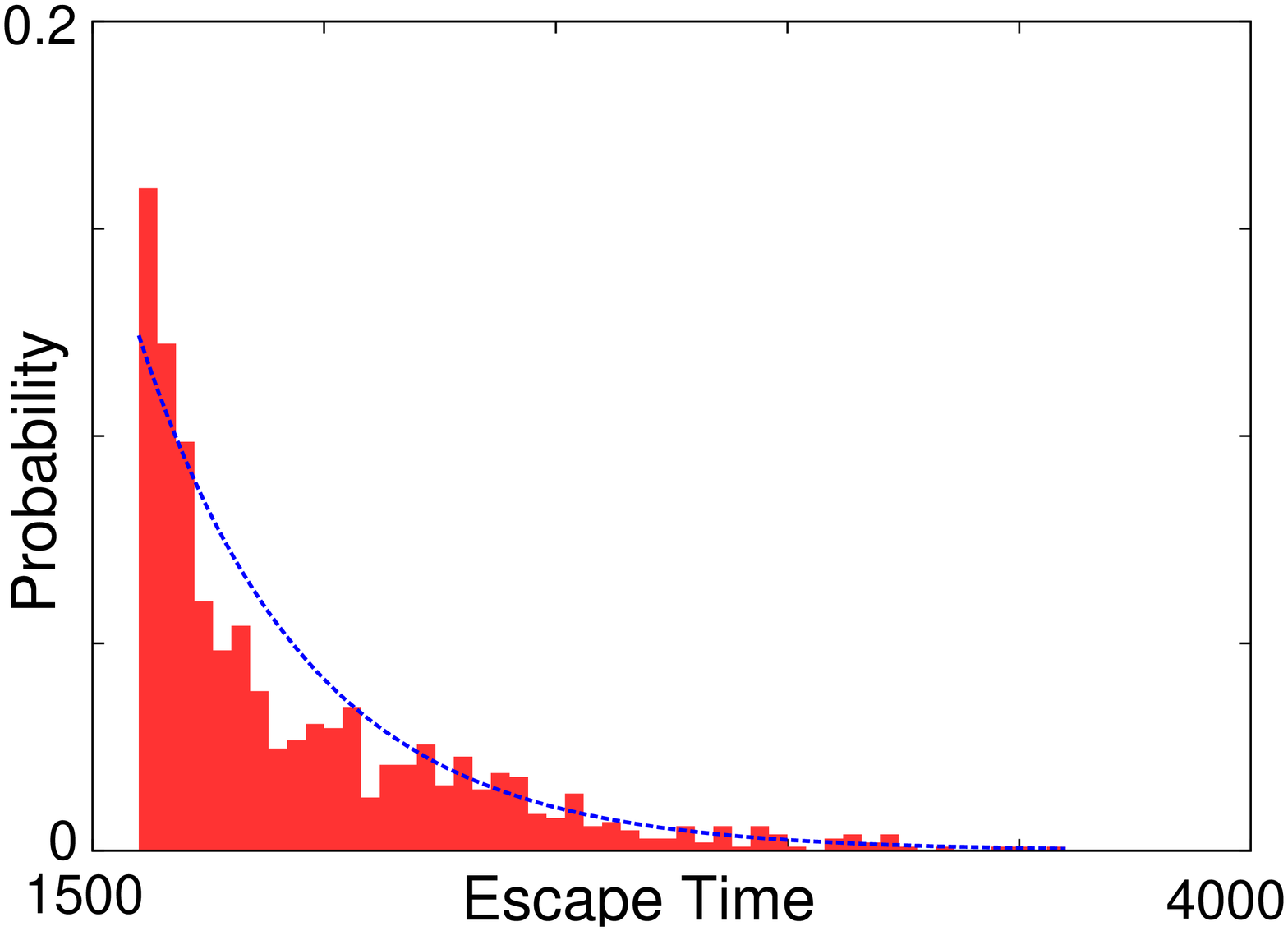}}\\
        \textbf{(c)}\resizebox{8cm}{!}{\includegraphics[angle=-90]{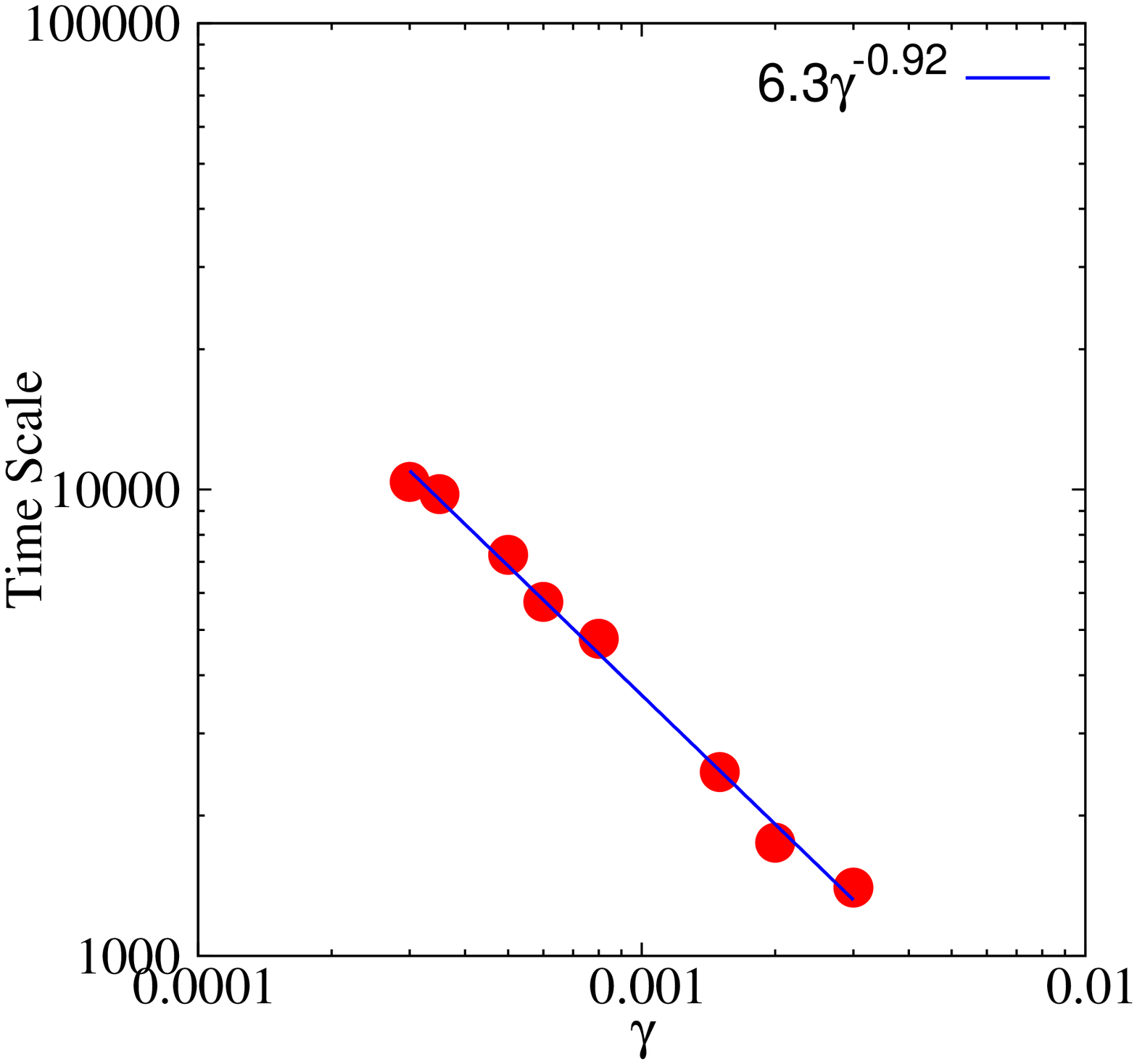}}

                \end{tabular}
 \caption{{(Color online) (a) Initial condition dependence of escape time, (b) distribution of escape times, with exponential fit $10.12e^{-0.00275x}$ and (c) fitted scaling of escape time scale with $\gamma$. In all the plots $\beta =0.1114,\gamma =0.002$ (except in (c) where it varies), $\lambda _{0}=-0.05,\phi _{ac}=0.01,\Omega =2\pi /5.9,dt=0.0002$; for initial conditions, refer to Appendix A. The escape time is calculated when the standard deviation of the consecutive 20 amplitudes of $\phi _{1},\phi _{2}$ and $\phi _{3}$ oscillations are all less than or equal to $0.005$.}}
\end{center}
\end{figure}

\begin{figure}[ht]
\begin{center}
\begin{tabular}{c}
        \textbf{(a)}\resizebox{8cm}{!}{\includegraphics[angle=0]{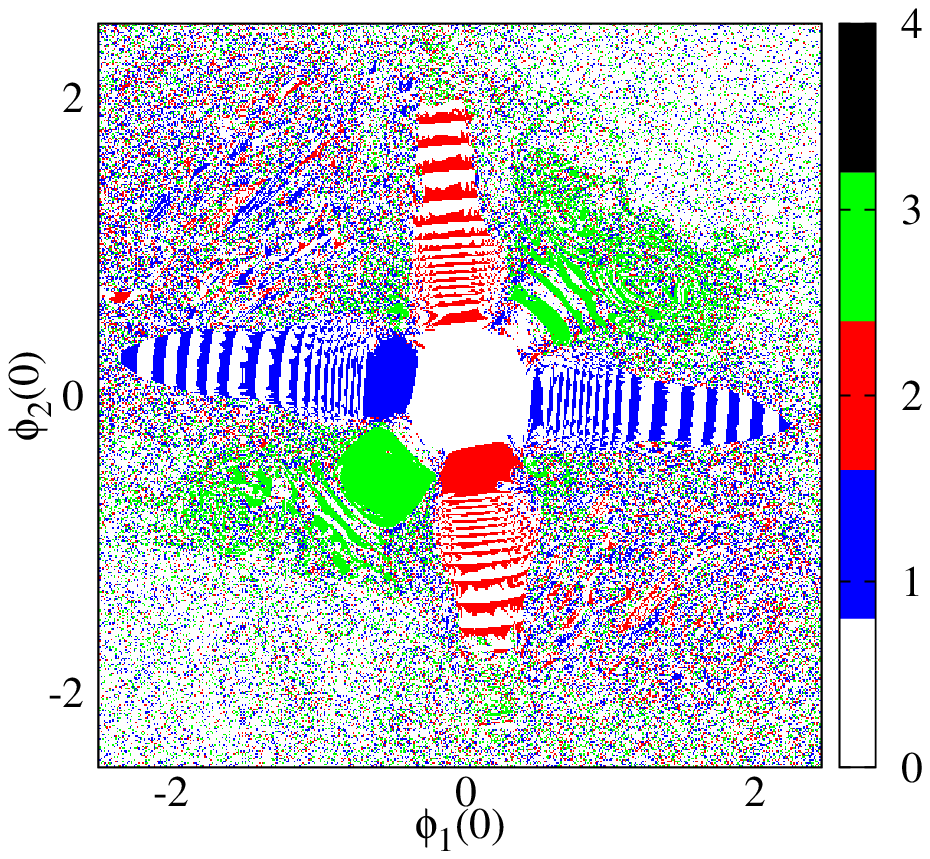}}
        \textbf{(b)}\resizebox{8cm}{!}{\includegraphics[angle=0]{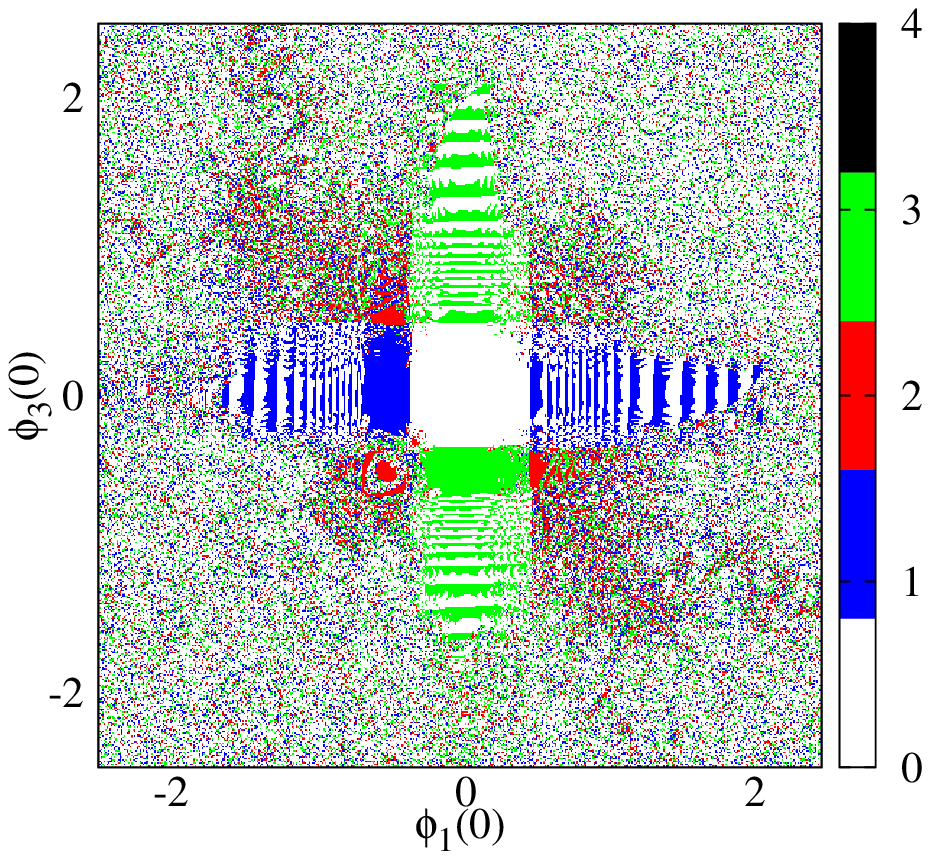}}\\
        \\
        \resizebox{7cm}{!}{\includegraphics[angle=-0]{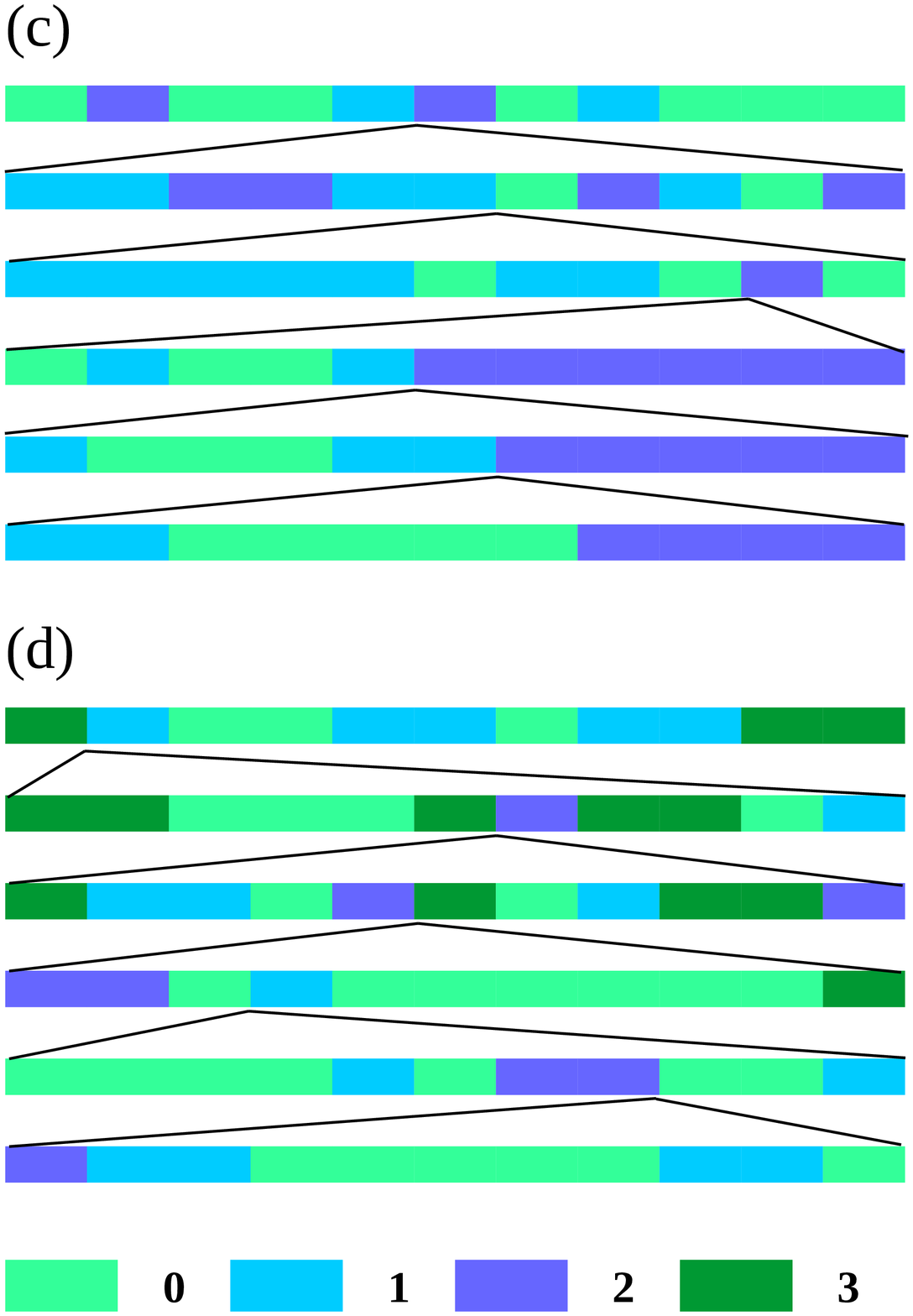}}

                  \end{tabular}
 \caption{{(Color online) Basins of Attraction for various dynamical states with $N=3,\beta =0.1114,\gamma =0.002,\lambda _{0}=-0.05,\phi _{ac}=0.01,\Omega =2\pi /5.9,dt=0.02,\dot\phi _{n}(0)=0$ for $n=1,2,3$ with (a) $\phi _{3}(0)=0$ and (b) $\phi _{2}(0)=0$. The number code reads: $0\Rightarrow$ synchronous; $1,2,3\Rightarrow$ chimera states with the number indicating the index of the non-phase-locked flux and $4\Rightarrow$ some other state (occurs only once in part a). Slices through the basin in (a) show the fractal and Wada nature: in each case, we initially choose 11 equally spaced initial conditions; then in the next step an adjacent pair of them is selected and including them as the endpoints again we choose 11 equally spaced initial conditions and repeat this to zoom basin boundaries with resolution increasing in each step. In (c) and (d), along the shown slices, $\phi _{1}(0)$ varies while $\phi _{2}(0)$ is fixed at $0.52$ and $1.0$ respectively and $dt=0.002$. The initial end points are $-0.4$ and $-0.3$ for (c) and $0.3$ and $0.4$ for (d).}}
\end{center}
\end{figure}


\newpage
\vskip 1cm
\begin{center}{\bf References}\end{center}

\begin{enumerate}

\bibitem{panaggio} M. J. Panaggio and D. M. Abrams, Nonlinearity 28 (3), R67 (2015).
\bibitem{pan1} ``Synchronization
From Coupled Systems to Complex Networks",
    S. Boccaletti, A. N. Pisarchik, C. I. del Genio, and
    A. Amann, Cambridge University Press, 1st edition (2018).

\bibitem{kur1} Y. Kuramoto and D. Battogtokh, Nonlinear Phenom. Complex Syst. 5, 380 (2002).

\bibitem{ma1a} R. Ma, J. Wang, and Z. Liu, Europhys. Lett., 91(4):40006 (2010).
\bibitem{ma1b}  T. A. Glaze1, S. Lewis and S. Bahar, Chaos 26, 083119 (2016).
\bibitem{ma1c} S. W. Haugland, L. Schmidt and K. Krischer, Nature Scientific Reports 5, Article number: 9883 (2015).

\bibitem{ral1a} R. G. Andrzejak, C. Rummel, F. Mormann, and K. Schindler, Sci. Rep. 6, 23000 (2016).

\bibitem{dwi1a} D. Barkley and L. S. Tuckerman, Phys. Rev. Lett. 94, 014502 (2005).

\bibitem{mar1a} E. A. Martensa, S. Thutupallic, A. Fourrièrec, and O. Hallatscheka, PNAS vol. 110 no. 26 10563-10567 (2013).
\bibitem{mar1b} J. Wojewoda, K. Czolczynski, Y. Maistrenko and T. Kapitaniak, Nature Scientific Reports 6, Article number: 34329 (2016).
\bibitem{mar1c} K. Blaha, R. J. Burrus, J. L. Orozco-Mora, E. Ruiz-Beltran, A. B. Siddique, V. D. Hatamipour and F. Sorrentino, Chaos 26, 116307 (2016).
\bibitem{mar1d} T. Kapitaniak, P. Kuzma, J. Wojewoda, K. Czolczynski and Y. Maistrenko, Nature Scientific Reports 4, Article number: 6379 (2014).
\bibitem{mar1e} T. Bountis, V. G. Kanas, J. Hizanidis, A. Bezerianos, Eur. Phys. J. Spec. Top. 223: 721 (2014).

\bibitem{zak1} M. Zaks and A. Pikovsky, Scientific Reports 7, 4648 (2017).

\bibitem{tin1a} M. R. Tinsley, S. Nkomo and K. Showalter, Nature Phys., 8:662–665 (2012).
\bibitem{tin1b} S. Nkomo, M. Tinsley and K. Showalter, Phys. Rev. Lett., 110:244102 (2013).
\bibitem{tin1c} M. R. Tinsley, S. Nkomo and K. Showalter, Chaos 26, 094826 (2016).
\bibitem{tin1d} L. Schmidt, K. Schonleber, K. Krischer and V. García-Morales, Chaos 24, 013102 (2014). 

\bibitem{dib1a} B. K. Bera, D. Ghosh, and M. Lakshmanan, Phys. Rev. E 93, 012205 (2016).
\bibitem{dib1b} B. K. Bera, D. Ghosh, and T. Banerjee, Phys. Rev. E 94, 012215 (2016).
\bibitem{dib1c} B. K. Bera and D. Ghosh, Phys. Rev. E 93, 052223 (2016). 
\bibitem{dib1d} M. I. Bolotov, G. V. Osipov, and A. Pikovsky, Phys. Rev. E 93, 032202 (2016).
\bibitem{dib1e} J. Hizanidis, V. G. Kanas, A. Bezerianos, and T. Bountis, Int. J. Bifurcation Chaos 24, 1450030 (2014).
\bibitem{dib1f} J. Hizanidis, N. E. Kouvaris, G. Zamora-Lopez, A. Diaz-Guilera and C. G. Antonopoulos, Scientific Reports 6, Article number: 19845 (2016).

\bibitem{hag1a} A. M. Hagerstrom,	T. E. Murphy,	R. Roy,	P. Hovel,	I. Omelchenko,	and E. Scholl, Nature Physics 8, 658–661(2012).
\bibitem{hag1b} N. Semenova, A. Zakharova, E. Scholl, and V. Anishchenko, EPL, 112, 4 (2015).
\bibitem{hag1c} A. V. Cano and M. G. Cosenza, Phys. Rev. E 95, 030202(R) (2017).

\bibitem{laz1a} N. Lazarides, G. Neofotistos, and G. P. Tsironis, Phys. Rev. B 91, 054303 (2015).
\bibitem{laz1b} J. Hizanidis, N. Lazarides, and G. P. Tsironis, Phys. Rev. E 94, 032219 (2016).
\bibitem{laz1c} J. Hizanidis, N. Lazarides, G. Neofotistos, and G.P. Tsironis, Eur. Phys. J. Special Topics 225, 1231–1243 (2016).

\bibitem{onew1} O. E. Omelchenko, Nonlinearity 31 R121 (2018).

\bibitem{wol1} M. Wolfrum and O. E. Omelchenko, Phys. Rev. E 84, 015201(R) (2011).
\bibitem{wol2} M. Wolfrum, O. E. Omelchenko, S. Yanchuk, and Y. L. Maistrenko, chaos 21, 013112 (2011).
\bibitem{olmi1} S. Olmi, E. A. Martens, S. Thutupalli, and A. Torcini, Phys. Rev. E 92, 030901(R) (2015).
\bibitem{omel1} O. E. Omelchenko, M. Wolfrum, and Y. L. Maistrenko, Phys. Rev. E 81, 065201(R) (2010).
\bibitem{dud1} D. Dudkowski, Y. Maistrenko, and T. Kapitaniak, Phys. Rev. E 90, 032920 (2014).

\bibitem{loo1} S. A. M. Loos, J. C. Claussen, E. Scholl, and A. Zakharova, Phys. Rev. E 93, 012209 (2016); R. G. Andrzejak, C. Rummel, F. Mormann, and K. Schindler, Sci. Rep. 6, 23000 (2016); D. P. Rosin, D. Rontani, N. D. Haynes, E. Scholl, and D. J. Gauthier, Phys. Rev. E 90, 030902(R) (2014); J. Sieber, O. E. Omelchenko, and M. Wolfrum, Phys. Rev. Lett. 112, 054102 (2014).

\bibitem{tel1} \textit{"Transient Chaos"}, Y.-C. Lai and Tamas Tel, Springer (2011).
\bibitem{tel2} Tamas Tel, Chaos, 25, 097619 (2015).

\bibitem{jan1} I. M. Janosi, L. Flepp, and Tamas Tel, Phys. Rev. Lett. 73, 529 (1994). 

\bibitem{aref1} H. Aref et. al., Rev. Mod. Phys. 89, 025007 (2017). 



\bibitem{jung2a} P. Jung1, A. V. Ustinov, and S. M. Anlage, Supercond. Sci. Technol. 27 073001 (2014).
\bibitem{jung2b} S. M. Anlage, J. Opt. 13, 024001 (2011).  
\bibitem{jung2c} P. Jung, S. Butz, M. Marthaler, M. V. Fistul, J. Leppakangas, V. P. Koshelets, and A. V. Ustinov, Nat. Commun. 5, 3730 (2014).
\bibitem{jung2d} M. Trepanier, Daimeng Zhang, Oleg Mukhanov, and Steven M. Anlage, Phys. Rev. X 3, 041029 (2013).


\bibitem{seth1} G. C. Sethia and A. Sen, Phys. Rev. Lett. 112, 144101 (2014); M. G. Clerc, S. Coulibaly, M. A. Ferre, M. A. Garcia-Ñustes, and R. G. Rojas, Phys. Rev. E 93, 052204 (2016); C. R. Laing, Phys. Rev. E 92, 050904(R) (2015); B.-W. Li and H. Dierckx, Phys. Rev. E 93, 020202(R) (2016).

\bibitem{zha1} D. Zhang, M. Trepanier, O. Mukhanov, and S. M. Anlage, Phys. Rev. X 5, 041045 (2015). 

\bibitem{dai1} D. Zhang, M. Trepanier, T. Antonsen, E. Ott, and S. M. Anlage, Phys. Rev. B 94, 174507 (2016); M. Trepanier, D. Zhang, O. Mukhanov, V. P. Koshelets, P. Jung, S. Butz, Ed. Ott, T. M. Antonsen, A. V. Ustinov, and S. M. Anlage, Phys. Rev. E 95, 050201(R) (2017).

\bibitem{likh1} K. K. Likharev, Dynamics of Josephson Junctions and Circuits (Gordon and Breach, Philadelphia, 1986).

\bibitem{ban1} A. Banerjee and M. Acharyya, Phys. Rev. E 94, 022213 (2016).

\bibitem{gopal1} R. Gopal, V. K. Chandrasekhar, A. Venkatesan, and M. Lakshmanan, Phys. Rev. E, 89, 052914 (2014).

\bibitem{gopal2} R. Gopal, V. K. Chandrasekhar, A. Venkatesan, and M. Lakshmanan, Phys. Rev. E, 91, 062916 (2015).

\bibitem{acebron1} J. A. Acebron, L. L. Bonilla, C. J. Perez Vicente, F. Ritort, and R. Spigler, Rev. Mod. Phys. 77, 137 (2005).

\bibitem{ash1} P. Ashwin, and O. Burylko, Chaos 25, 013106 (2015).

\bibitem{ash2} C. Bick, and P. Ashwin, Nonlinearity 29, 1468–1486 (2016).

\bibitem{yur1} Y. Maistrenko, S. Brezetsky, P. Jaros, R. Levchenko, and T. Kapitaniak, Phys. Rev. E 95, 010203(R) (2017).

\bibitem{and2} A. Rohm, F. Bohm, and K. Ludge, Phys. Rev. E 94, 042204 (2016).

\bibitem{wo1} J. Wojewoda, K. Czolczynski, Y. Maistrenko, and T. Kapitaniak, Scientific Reports 6, 34329 (2016). 

\bibitem{rajar1} J. D. Hart, K. Bansal, T. E. Murphy, and R. Roy
Chaos 26 (9), 094801 (2016).

\bibitem{chaos1} E. Ott, Chaos in Dynamical Systems, 2nd ed. (Cambridge University Press, Cambridge, England, 2002); T. Tel and M. Gruiz, Chaotic Dynamics: An Introduction Based on Classical Mechanics (Cambridge University Press, Cambridge, England, 2006).
 
\bibitem{agu1} J. Aguirre, R. L. Viana, and M. A. F. Sanjuan, Rev. Mod. Phys. 81, 333 (2009).

\bibitem{xiao1} X. Chen, T. Nishikawa, and A. E. Motter, Phys. Rev. X 7, 021040 (2017).

\bibitem{adil1} A. E. Motter, M. Gruiz, G. Karolyi, and T. Tel, Phys. Rev. Lett. 111, 194101 (2013).


\bibitem{ble1} F. Blesa, J. M. Seoane, R. Barrio, and M. A. F. Sanjuan, Phys. Rev. E 89, 042909 (2014).

\bibitem{alv1} A. Daza, A. Wagemakers, M. A. F. Sanjuan, and J. A. Yorke, Scientific Reports 5, Article number: 16579 (2015).


\bibitem{gra1} C. Granata, A. Vettoliere, R. Russo, M. Russo, and B. Ruggiero, Phys. Rev. B 83, 092504 (2011). 

\bibitem{lan1} T. Lanting, M. H. Amin, A. J. Berkley, C. Rich, S.-F. Chen, S. LaForest, 3 and R de Sousa, Phys. Rev. B 89, 014503 (2014).

\bibitem{bial1} R. C. Bialczak, R. McDermott, M. Ansmann, M. Hofheinz, N. Katz, E. Lucero, M. Neeley, A. D. O’Connell, H. Wang, A. N. Cleland, and J. M. Martinis, Phys. Rev. Lett. 99, 187006 (2007); R. H. Koch, D. P. DiVincenzo, and J. Clarke, Phys. Rev. Lett. 98, 267003 (2007); S. Sendelbach, D. Hover, A. Kittel, M. Muck, J. M. Martinis, and R. McDermott, Phys. Rev. Lett. 100, 227006.
(2008).

\bibitem{pala1} E. Paladino, Y. M. Galperin, G. Falci, and B. L. Altshuler, Rev. Mod. Phys. 86, 361 (2014).
\end{enumerate}
\end{document}